\documentclass[12pt,journal,draftcls,a4paper,onecolumn]{IEEEtran}

\usepackage{dsfont}
\usepackage{amsmath}
\usepackage{amssymb}
\usepackage{amsfonts}
\usepackage{graphicx}
\usepackage{amsmath}
\usepackage[noadjust]{cite}
\usepackage[all,poly]{xy}
\usepackage{multirow}

\newcommand{\nosou}{m}
\newcommand{\nbsou}{M}
\newcommand{\noband}{j}
\newcommand{\nbband}{L}
\newcommand{\noobs}{i}
\newcommand{\nbobs}{N}
\newcommand{\MATobs}{\mathbf Y}
\newcommand{\Vobs}[1]{{\mathbf y}_{#1}}
\newcommand{\obs}[2]{y_{#1,#2}}

\newcommand{\MATsou}{\mathbf S}
\newcommand{\Vsou}[1]{{\mathbf s}_{#1}}
\newcommand{\bfs}{\mathbf s}

\newcommand{\sou}[2]{s_{#1,#2}}
\newcommand{\abond}[2]{c_{#1,#2}}
\newcommand{\Vabond}[1]{{\bold c}_{#1}}
\newcommand{\MATabond}{\mathbf C}
\newcommand{\mix}[2]{a_{#1,#2}}
\newcommand{\Vmix}[1]{{\bold a}_{#1}}
\newcommand{\MATmix}{\mathbf A}
\newcommand{\Vnoise}[1]{{\mathbf e}_{#1}}
\newcommand{\noise}[2]{e_{#1,#2}}
\newcommand{\MATnoise}{\mathbf E}
\newcommand{\Vnoisevar}{{\boldsymbol \sigma}_{\mathrm{e}}^2}
\newcommand{\noisevar}[1]{\sigma_{\mathrm{e},#1}^2}

\newcommand{\Vsouvar}{{\boldsymbol \sigma}_{\mathrm{s}}^2}
\newcommand{\souvar}[1]{\sigma_{\mathrm{s},#1}^2}
\newcommand{\soucoeffA}[1]{\alpha_{#1}}
\newcommand{\soucoeffB}[1]{\beta_{#1}}
\newcommand{\VsoucoeffA}{\boldsymbol{\alpha}}
\newcommand{\VsoucoeffB}{\boldsymbol{\beta}}

\newcommand{\transp}{^{\mathsf T}}
\newcommand{\inv}{^{-1}}
\newcommand{\R}{\mathbb{R}}
\newcommand{\paramvect}{\boldsymbol{\Theta}}
\newcommand{\hypervect}{\boldsymbol{\Phi}}

\newcommand{\un}{{\bf u}}
\newcommand{\Simplex}{\mathbb{S}}

\newcommand{\Id}{\textbf{I}}
\newcommand{\Indicfun}[2]{\textbf{1}_{#1}(#2)}
\newcommand{\sampleVsou}[2]{\widetilde{\Vsou{}}_{#1}^{(#2)}}

\newcommand{\sampleVmix}[2]{\widetilde{\Vmix{}}^{(#2)}_{#1}}

\newcommand{\samplenoisevar}[2]{{\left\{\widetilde{\sigma}_{\mathrm{e},#1}^{2}\right\}^{(#2)}}}
\newcommand{\samplehypernoise}[1]{\widetilde{\psi}_{\mathrm{e}}^{(#1)}}
\newcommand{\samplesoucoeffA}[2]{\widetilde{\alpha}_{#1}^{(#2)}}
\newcommand{\samplesoucoeffB}[2]{\widetilde{\beta}_{#1}^{(#2)}}

\newenvironment{algogo}[1]{
\bigskip \noindent 
\bigskip
\par \refstepcounter{algo} {\bf \textsc{Algorithm \thealgo.}} #1 } {
\smallskip }
\newcounter{algo}
\renewcommand{\thealgo}{\arabic{algo}}

\newcommand{\halffigwidth}{0.4\columnwidth}

\title{Bayesian separation of spectral sources under \\non-negativity and full additivity constraints}

\author{Nicolas Dobigeon$^{1,2}$, Sa\"id Moussaoui$^{3}$, \\Jean-Yves Tourneret$^{1}$ and
C\'edric Carteret$^{4}$
\\
\normalsize $^1$ University of Toulouse, IRIT/INP-ENSEEIHT, 2 rue Camichel, 31071 Toulouse, France. \\
\normalsize $^2$ University of Michigan, Department of EECS, Ann Arbor, MI 48109-2122, USA \\
\normalsize $^3$ IRCCyN/ECN, 1 rue de la No\"e, BP 92101, 44321 Nantes cedex 3, France \\
\normalsize $^4$ University of Nancy, LCPME, 405 rue de Vandoeuvre, 54600 Villers-l\`es-Nancy, France \\
\small\texttt{\{Nicolas.Dobigeon,Jean-Yves.Tourneret\}@enseeiht.fr,\\
Said.Moussaoui@irccyn.ec-nantes.fr,
Cedric.Carteret@lcpme.cnrs-nancy.fr}}

\begin{document}
\maketitle

\begin{abstract}
This paper addresses the problem of separating spectral sources
which are linearly mixed with unknown proportions. The main
difficulty of the problem is to ensure the full additivity
(sum-to-one) of the mixing coefficients and non-negativity of
sources and mixing coefficients. A Bayesian estimation approach
based on Gamma priors was recently proposed to handle the
non-negativity constraints in a linear mixture model. However,
incorporating the full additivity constraint requires further
developments. This paper studies a new hierarchical Bayesian model
appropriate to the non-negativity and sum-to-one constraints
associated to the sources and the mixing coefficients of linear
mixtures. The estimation of the unknown parameters of this model is
performed using samples obtained with an appropriate Gibbs
algorithm. The performance of the proposed algorithm is evaluated
through simulation results conducted on synthetic mixture data. The
proposed approach is also applied to the processing of
multicomponent chemical mixtures resulting from Raman spectroscopy.
\end{abstract}


\newpage
\section{Introduction} \label{sec:intro}
Blind source separation (BSS) is a signal processing problem arising
in many applications where one is interested by extracting signals
that are observed as mixtures \cite{Cichocki2002}. Pioneering works
dealing with this problem have focused on the mutual statistical
independence of the sources, which led to the well known independent
component analysis (ICA)
\cite{Comon1991,Comon1994,Cardoso1998,Hyvarinen2001}. However, when
the sources and the mixing coefficients have to satisfy specific
constraints the resulting constrained source separation problem
becomes more complicated. Therefore appropriate separation
algorithms have to be developed to handle these constraints. When
the sources are actually independent, ICA provides estimates of the
sources and mixing coefficients which implicitly satisfy these
constraints. However, these algorithms, that try to maximize the
independence between the estimated sources, have not been designed
for correlated sources.

Non-negativity is a physical constraint which has retained a growing
attention during the last decade. For instance, Plumbley and his
co-authors have addressed the case of non-negative independent
sources and proposed the non-negative independent component analysis
algorithm \cite{Plumbley2003}. The case of both non-negative sources
and non-negative mixing coefficients has been handled by using
non-negative matrix factorization algorithms (NMF) \cite{Sajda2004}
and a Bayesian positive source separation algorithm
\cite{Moussaoui2006b}. By adding a source sparsity constraint, a
method ensuring the sparseness of the sources (referred to as
non-negative sparse coding) has been presented in \cite{Hoyer2004}.
A Bayesian approach allowing one to perform the separation of sparse
sources has also been proposed in \cite{Fevotte2006} using a
T-student distribution. Cauchy Hyperbolic priors have been
introduced in \cite{Snoussi2006} without considering the
non-negativity constraint.

This paper addresses a source separation problem in the case of
linear instantaneous mixtures where the source signals are
non-negative and the mixing coefficients satisfy non-negativity and
full additivity constraints. These constraints have been observed in
many applications. These applications include analytical chemistry
for the analysis of kinetic reactions monitored by spectroscopy
\cite{Malinowski2002} or image processing for the analysis of
hyperspectral images \cite{Chang2007}. A Bayesian framework
appropriate to constrained source separation problem is first
proposed. Prior distributions encoding non-negativity and full
additivity constraints are assigned to the source signals and mixing
coefficients. However, the standard Bayesian estimators resulting
from these priors have no simple closed form expression. As a
consequence, Markov chain Monte Carlo (MCMC) methods are proposed to
generate samples according to the full posterior distribution of the
unknown parameters. Estimators of the mixing coefficients and the
source signals are then constructed from these generated samples.
The paper is organized as follows. Section~\ref{sec:model} defines a
hierarchical Bayesian model (HBM) for the addressed constrained
source separation problem. In particular, prior distributions are
introduced such that they are concentrated on a simplex and they
satisfy the positivity and full additivity constraints.
Section~\ref{sec:Gibbs} describes a Gibbs sampling strategy that
allows one to overcome the computational complexity inherent to this
HBM. Simulations conducted on synthetic mixture data are presented
in Section~\ref{sec:simus_synth}. As a consequence, the performance
of the proposed Bayesian estimation algorithm can be appreciated for
constrained source separation problems. The interest of the proposed
Bayesian approach is also illustrated by the analysis of real
experimental data reported in Section~\ref{sec:simus_real}.
Conclusions and perspectives are reported in
Section~\ref{sec:conclusions}.

\section{Problem statement} \label{sec:problem}
The linear mixing model studied in this paper assumes that the
observed signal is a weighted sum of $\nbsou$ unknown sources. In
the case of spectral mixture data this model can be expressed by:
 \begin{equation} \label{eq:mixmodel}
 \obs{\noobs}{\noband} = \sum_{\nosou=1}^\nbsou \abond{\noobs}{\nosou}
 \sou{\nosou}{\noband}  +\noise{\noobs}{\noband},
\end{equation}
where $\obs{\noobs}{\noband}$ is the observed spectrum at
time/spatial index $\noobs$ ($\noobs=1,\ldots,\nbobs$) in the
$\noband^{\text{th}}$ spectral band ($\noband=1,\ldots,\nbband$),
$\nbobs$ is the number of observed spectra, $\nbsou$ is the number
of mixture components and $\nbband$ is the number of spectral bands.
The coefficient $\abond{\noobs}{\nosou}$ is the contribution of the
$\nosou^{\text{th}}$ component in the $\noobs^{\text{th}}$ mixture
and $\noise{\noobs}{\noband}$ is an additive noise modeling
measurement errors and model uncertainties. The linear mixing model
can be represented by the following matrix formulation:
\begin{equation}  \label{eq:mixmodelmat}%
  \MATobs = \MATabond \MATsou + \MATnoise,
\end{equation}
where the matrices $\MATobs =
\left[\obs{\noobs}{\noband}\right]_{\noobs,\noband} \in \R^{\nbobs
\times \nbband}$, $\MATabond  = \left[\abond{\noobs}{\nosou}
\right]_{\noobs,\nosou} \in \R^{\nbobs \times \nbsou}$, $\MATsou =
\left[\sou{\nosou}{\noband}\right]_{\nosou,\noband} \in \R^{\nbsou
\times \nbband}$ and $\MATnoise =
  \left[\noise{\noobs}{\noband}\right]_{\noobs,\noband} \in \R^{\nbobs \times
\nbband}$ contain respectively the observed spectra, the mixing
coefficients, the spectral sources and the additive noise
components. The noise sequences
$\Vnoise{\noobs}=\left[\noise{\noobs}{1},\ldots,\noise{\noobs}{\nbband}\right]\transp$
($\noobs=1,\ldots,\nbobs$) are assumed to be independent and
identically distributed (i.i.d.) according to zero-mean Gaussian
distributions with covariance matrices $\noisevar{\noobs}
\Id_{\nbband}$, where $\Id_{\nbband}$ is the $\nbband \times
\nbband$ identity matrix. Note that this last assumption implies
that the noise variances are the same in all the spectral bands.
This reasonable assumption has been considered in many recent works
including \cite{Moussaoui2006b} and \cite{Snoussi2006}. It could be
relaxed at the price of increasing the computational complexity of
the proposed algorithm \cite{Dobigeon2008icassp}.

In the framework of spectral data analysis, it is obvious from
physical considerations that both the mixing coefficients and the
source signals satisfy the following non-negativity constraints:
\begin{equation}\label{eq:positivityconstraint}
      \sou{\nosou}{\noband} \geqslant 0 \; \text{ and } \abond{\noobs}{\nosou}\geqslant 0,  \quad \forall
     (\noobs,\nosou,\noband).
\end{equation}
Moreover, in many applications, the mixing coefficients have also to
satisfy the full additivity constraint\footnote{This condition is
also referred to as \emph{sum-to-one} constraint in the
literature.}:
\begin{equation}\label{eq:additivityconstraint}
       \sum_{\nosou=1}^\nbsou \abond{\noobs}{\nosou} =1 \quad \forall \noobs.
\end{equation}
These applications include spectroscopy for the analysis of kinetic
reactions \cite{DeJuan2000} and hyperspectral imagery where the
mixing coefficients correspond to abundance fractions
\cite{Dobigeon2008}.

The separation problem addressed in this paper consists of jointly
estimating the abundances and the spectral sources under the
non-negativity and the full additivity constraints. There are
several methods allowing one to address the estimation problem under
non-negativity constraint. These methods include NMF methods
\cite{LeeSeung1999} and its variants \cite{Cichocki2002}. From a
Bayesian point of view an original model was proposed in
\cite{Moussaoui2006b} where Gamma priors are used to encode the
positivity of both the sources and the mixing coefficients. This
paper goes a step further by including the additivity of the mixing
coefficients in the Bayesian model. Note that this constraint allows
one to resolve the scale indeterminacy inherent to the linear mixing
model even if non-negativity constraint is imposed. Indeed, this
full additivity constraint enforces the $\ell_1$ norm of each
concentration vector $\Vabond{\noobs}$ to be equal to
$\left\|\Vabond{\noobs}\right\|_1 = \sum_{m=1}^M |c_{i,m}|= 1$.

\section{Hierarchical Bayesian Model} \label{sec:model}
The unknown parameter vector for the source separation problem
described previously is $\paramvect = \left( \MATsou, \MATabond,
\boldsymbol{\sigma}^2_e \right)$ where  $\MATsou$ and $\MATabond$
are the source and concentration matrices and
$\boldsymbol{\sigma}^2_e =(\sigma_{e,1},\ldots,\sigma_{e,N})^T$
contains the noise variances. Following the Bayesian estimation
theory, the inference of the unknown parameters from the available
data $\MATobs$ is based on the posterior distribution
$f\left(\paramvect|\MATobs\right)$, which is related to the
observation likelihood $f\left(\MATobs|\paramvect\right)$ and the
parameter priors $f\left(\paramvect\right)$ via the Bayes' theorem:
$$
f\left(\paramvect|\MATobs\right)\propto
f\left(\MATobs|\paramvect\right) \;  f\left(\paramvect\right),
$$ where $\propto$ means ``proportional to". The observation likelihood
and the parameters priors are detailed in the sequel.

\subsection{Observation likelihood}
The statistical assumptions on the noise vector $\Vnoise{\noobs}$
and the linear mixing model described in \eqref{eq:mixmodel} allow
one to write:
\begin{equation}
\Vobs{\noobs} \mid \MATsou,  \Vabond{\noobs}, \noisevar{\noobs}\sim
\mathcal{N}\left(\MATsou\transp\Vabond{\noobs},\noisevar{\noobs}\Id_{\nbband}\right),
\end{equation}
where
$\Vobs{\noobs}=\left[\obs{\noobs}{1},\ldots,\obs{\noobs}{\nbband}\right]\transp$,
$\Vabond{\noobs}=\left[\abond{\noobs}{1},\ldots,\abond{\noobs}{\nbsou}\right]\transp$
and $\mathcal{N}(\cdot,\cdot)$ denotes the Gaussian distribution. By
assuming the mutual independence between the vectors
$\Vnoise{1},\ldots,\Vnoise{\nbobs}$, the likelihood of $\MATobs$ is:
\begin{equation}
\label{eq:likelihood}
  f\left(\MATobs|\MATabond,\MATsou,\Vnoisevar\right) \propto  \frac{1}{
\prod_{\noobs=1}^{\nbobs} \sigma_{\mathrm{e},\noobs}^{\nbband}}
\exp\left( - \sum_{\noobs=1}^{\nbobs}
\frac{\left\|\Vobs{\noobs}-\MATsou\transp\Vabond{\noobs}\right\|_2^2}{2\noisevar{\noobs}}\right),
\end{equation}
where $\left\|\mathbf{x} \right\|_2= \left(\mathbf{x}\transp
\mathbf{x}\right)^{\frac{1}{2}}$ stands for the standard $\ell_2$
norm.

\subsection{Parameter Priors}
\subsubsection{Concentrations}
In order to ensure the non-negativity and additivity constraints,
the concentrations are assigned a Dirichlet prior distribution. This
distribution is frequently used in statistical inference for
positive variables summing to one. The Dirichlet probability density
function (pdf) is defined by:
\begin{equation}
\mathcal{D}(\Vabond{\noobs} | \delta_1, \ldots, \delta_\nbsou) =
\frac{\Gamma\left(\sum_{\nosou=1}^{\nbsou} \delta_\nosou\right)}{
\prod_{\nosou=1}^{\nbsou}\Gamma (\delta_\nosou) } \left(
\prod_{\nosou=1}^{\nbsou} c^{\delta_\nosou-1}_{\noobs,\nosou}
\right) \Indicfun{\left\{c_{\noobs,\nosou}\geqslant 0 ; \,
\sum_{\nosou=1}^{\nbsou} c_{\noobs,\nosou}=1\right\}
}{\Vabond{\noobs}},
\end{equation}
where $\delta_1, \ldots,\delta_\nbsou
0$ are the Dirichlet distribution parameters,
$\Gamma\left(\cdot\right)$ is the Gamma function and
$\Indicfun{A}{.}$ denotes the indicator function defined on the set
$A$:
\begin{equation}
  \left\{
\begin{array}{ll}
    \Indicfun{A}{\mathbf{x}}=1, & \hbox{if $\mathbf{x} \in A$;} \\
    \Indicfun{A}{\mathbf{x}}=0, & \hbox{otherwise.} \\
\end{array}
\right.
\end{equation}
According to this prior, the expected value of the $m^\text{th}$
spectral source abundance is $\mathbb{E}[c_{\noobs,\nosou}] =
\delta_\nosou /\sum\limits_{\nosou=1}^{\nbsou}  \delta_\nosou$. We
assume here that the abundances are \textit{a priori} equiprobable
(reflecting the absence of knowledge regarding these parameters)
which corresponds to identical parameters $\left\{\delta_\nosou=1,
\forall \nosou =1,\ldots,M \right\}$. An interesting
reparametrization can be introduced here to handle the full
additivity constraint. This reparametrization consists of splitting
the concentration vectors into two parts\footnote{From a practical
point of view, it is interesting to note that the component of
$\Vmix{\noobs}$ to be discarded will be randomly chosen at each
iteration of the Algorithm introduced in Section~\ref{sec:Gibbs}.}:
\begin{equation}
\Vabond{\noobs} =
\left[\Vmix{\noobs}\transp,\abond{\noobs}{\nbsou}\right] \transp,
\end{equation}
where $\Vmix{\noobs}\transp =
\left[\abond{\noobs}{1},\ldots,\abond{\noobs}{\nbsou-1}\right] $ and
$ \abond{\noobs}{\nbsou} =
1-\sum_{\nosou=1}^{\nbsou-1}\abond{\noobs}{\nosou}$. It induces a
new unknown parameter vector
$\paramvect=\left\{\MATmix,\MATsou,\Vnoisevar\right\}$ (the same
notation is used for this new parameter vector to avoid defining new
variables). The proposed prior for $\Vmix{\noobs}$, $\noobs =
1,\ldots,\nbobs$ is a uniform distribution on the following simplex:
\begin{equation}
\label{eq:space_S}
  \Simplex=\left\{ \Vmix{\noobs} ;  \mix{\noobs}{\nosou} \geq 0, \ \forall \nosou =1,\ldots,\nbsou-1,
\; \sum_{\nosou=1}^{\nbsou-1} \mix{\noobs}{\nosou} \leq 1 \right\}.
\end{equation}
By assuming \textit{a priori} mutual independence between the
vectors $\Vmix{\noobs}$, the prior distribution for the matrix
$\MATmix = \left[\Vmix{1},\ldots,\Vmix{\nbobs}\right]\transp$
reduces to:
\begin{equation}
\label{eq:abond_prior}
  f\left(\MATmix\right) \propto \prod_{\noobs=1}^{\nbobs}
  \Indicfun{\Simplex}{\Vmix{\noobs}}.
\end{equation}

\subsubsection{Source signals}  \label{subsubsec:sources_prior}
To take into account the non-negativity constraint, the two
parameter Gamma distribution seems to be a good candidate thanks to
its flexibility, i.e. the pdf has many different shapes depending on
the values of its parameters (see \cite{Moussaoui2006b} for
motivations). This distribution encodes positivity and covers a wide
range of distribution shapes\footnote{A more general model would
consist of using a mixture of Gamma distributions as in
\cite{Hsiao2003}. However, the Gamma distribution which leads to a
simple Bayesian model has been preferred here for simplicity.}. The
assumption of independent source samples leads to a prior
distribution for each spectral source expressed as:
\begin{equation}\label{eq:Vsou_prior}
f\left(\Vsou{\nosou}\big|\soucoeffA{\nosou},\soucoeffB{\nosou}\right)
=
\left[\frac{\soucoeffB{\nosou}^{\soucoeffA{\nosou}}}{\Gamma\left(\soucoeffA{\nosou}\right)}\right]^\nbband
\prod_{\noband=1}^\nbband \left[
\sou{\nosou}{\noband}^{\soucoeffA{\nosou}-1}
\exp\left(-\soucoeffB{\nosou}\sou{\nosou}{\noband}\right)
\Indicfun{\mathbb{R}^+}{\sou{\nosou}{\noband}}\right].
\end{equation}
Note that this distribution generalizes the exponential prior
presented in \cite{Miskin2001, Dobigeon2007ssp} and has the
advantage of providing a wider variety of distributions (see also
paragraph~\ref{subsec:modified_BM} for additional details regarding
the exponential prior). Finally, by assuming the mutual independence
between the spectral sources, we obtain the following prior
distribution for $\MATsou$:
\begin{equation}
\label{eq:mat_prior}
  f\left(\MATsou\big|\VsoucoeffA, \VsoucoeffB\right) = \prod_{\nosou=1}^\nbsou
  f\left(\Vsou{\nosou} \big| \soucoeffA{\nosou},\soucoeffB{\nosou}\right),
\end{equation}
where
$\VsoucoeffA=\left[\soucoeffA{1},\ldots,\soucoeffA{\nbsou}\right]\transp$
and
$\VsoucoeffB=\left[\soucoeffB{1},\ldots,\soucoeffB{\nbsou}\right]\transp$
are the source hyperparameter vectors.

\subsubsection{Noise variances}
Conjugate priors which are here inverse Gamma (IG) distributions are
chosen for each noise variance $\noisevar{\noobs}$ \cite[App.
A]{Robert2001}:
\begin{equation}
\label{eq:var_prior_ind}
\noisevar{\noobs}\big|\rho_{\mathrm{e}},\psi_{\mathrm{e}} \sim
\mathcal{IG}\left(\frac{\rho_{\mathrm{e}}}{2},\frac{\psi_{\mathrm{e}}}{2}\right),
\end{equation}
where $\mathcal{IG}\left(a,b\right)$ denotes the IG distribution
with parameters $a$ and $b$. Note that choosing conjugate
distributions as priors makes the Bayesian analysis easier
\cite[Chap. 2]{Bishop2006}. By assuming the independence between the
noise variances $\noisevar{\noobs}$, $\noobs =1,\ldots,\nbobs$, the
prior distribution of $\Vnoisevar$ is:
\begin{equation}
\label{eq:var_prior}
f\left(\Vnoisevar\big|\rho_{\mathrm{e}},\psi_{\mathrm{e}}\right) =
\prod_{\noobs=1}^\nbobs
f\left(\noisevar{\noobs}\big|\rho_{\mathrm{e}},\psi_{\mathrm{e}}\right).
\end{equation}
The hyperparameter $\rho_{\mathrm{e}}$ will be fixed to
$\rho_{\mathrm{e}}=2$ whereas $\psi_{\mathrm{e}}$ is an adjustable
hyperparameter as in \cite{Punskaya2002}.

\subsection{Hyperparameter priors}

The hyperparameter vector associated with the prior distributions
previously introduced is
$\hypervect=\left\{\VsoucoeffA,\VsoucoeffB,\psi_{\mathrm{e}}\right\}$.
Obviously, the BSS performances depend on the values of these
hyperparameters. In this paper, we propose to estimate them within a
fully Bayesian framework by assigning them non-informative prior
distributions. This naturally introduces a second level of hierarchy
within the Bayes' paradigm, resulting in a so-called hierarchical
Bayesian model \cite[p. 299]{Robert1999}.

\subsubsection{Source hyperparameters}
Conjugate exponential densities with parameters
$\lambda_{\soucoeffA{\nosou}}$ have been chosen as prior
distributions for the hyperparameters $\soucoeffA{\nosou}$
\cite[App. A]{Robert2001}:
\begin{equation}\label{eq:soucoeffA_prior}
  \soucoeffA{\nosou} \big|
  \lambda_{\soucoeffA{\nosou}} \sim \mathcal{E}\left(\lambda_{\soucoeffA{\nosou}}\right).
\end{equation}
Conjugate Gamma distributions with parameters
$\left(\alpha_{\soucoeffB{\nosou}},\beta_{\soucoeffB{\nosou}}\right)$
have been elected as prior distributions for the hyperparameters
$\soucoeffB{\nosou}$ \cite[App. A]{Robert2001}:
\begin{equation}\label{eq:soucoeffB_prior}
  \soucoeffB{\nosou}\big|
  \alpha_{\soucoeffB{\nosou}},\beta_{\soucoeffB{\nosou}} \sim
  \mathcal{G}\left(\alpha_{\soucoeffB{\nosou}},\beta_{\soucoeffB{\nosou}}\right).
\end{equation}
The fixed hyperparameters $\left\{\alpha_{\soucoeffB{\nosou}},
\beta_{\soucoeffB{\nosou}},
\lambda_{\soucoeffA{\nosou}}\right\}_\nosou$ have been chosen to
obtain flat priors, i.e. with large variances: $\alpha_{\beta_m} =2,
\beta_{\beta_m}=10^{-2}$ and $\lambda_{\soucoeffA{\nosou}}=10^{-2}$.

\subsubsection{Noise variance hyperparameters}
The prior for $\psi_{\mathrm{e}}$ is a non-informative Jeffreys'
prior which reflects the lack of knowledge regarding this
hyperparameter:
\begin{equation}
\label{eq:hypernoise_var_prior} f\left(\psi_{\mathrm{e}}\right)
\propto
\frac{1}{\psi_{\mathrm{e}}}\Indicfun{\R^+}{\psi_{\mathrm{e}}}.
\end{equation}
Assuming the independence between the hyperparameters, the prior
distribution of the hyperparameter vector
$\hypervect=\left\{\VsoucoeffA,\VsoucoeffB,\psi_{\mathrm{e}}\right\}$
can be written as:
\begin{equation}
\begin{split}
\label{eq:hyper_prior} f\left(\hypervect\right) &\propto
\prod_{\nosou=1}^\nbsou\left[\lambda_{\soucoeffA{\nosou}}
                        \exp\left(-\lambda_{\soucoeffA{\nosou}}\soucoeffA{\nosou}\right)
                        \Indicfun{\mathbb{R}^+}{\soucoeffA{\nosou}}\right]\\
                        &\times \prod_{\nosou=1}^\nbsou\left[
                        \soucoeffB{\nosou}^{\alpha_{\soucoeffB{\nosou}}-1}
                        \exp\left(-\beta_{\soucoeffB{\nosou}}\soucoeffB{\nosou}\right)
                        \Indicfun{\mathbb{R}^+}{\soucoeffB{\nosou}}
                        \right] \frac{1}{\psi_{\mathrm{e}}}\Indicfun{\R^+}{\psi_{\mathrm{e}}}.
\end{split}
\end{equation}

\subsection{Posterior distribution of $\paramvect$}\label{subsec:posterior}
\begin{figure*}
\centering{
\includegraphics[width=0.9\textwidth]{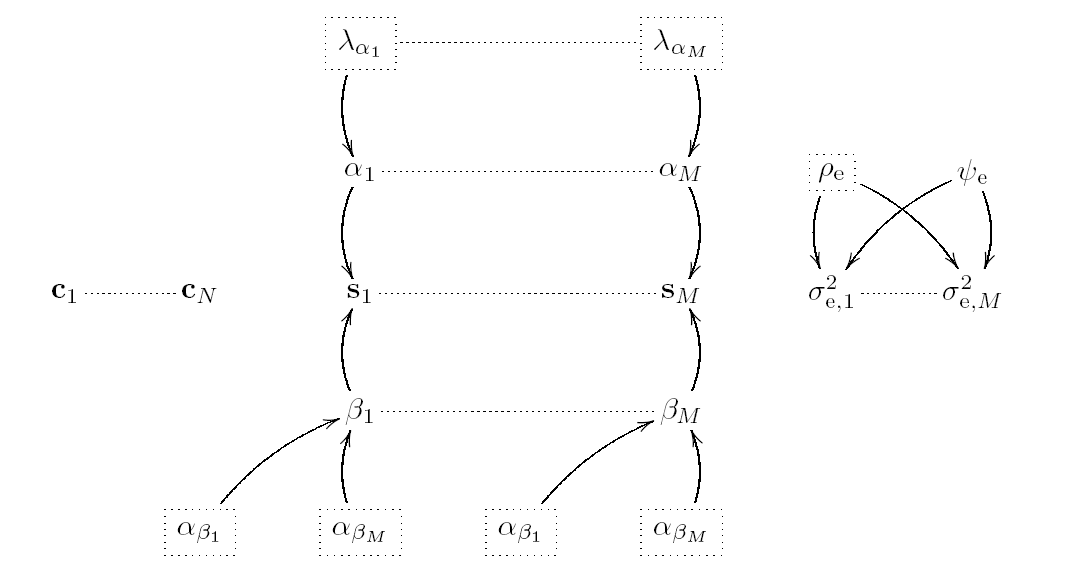}
} \caption{DAG for the parameter priors and hyperpriors (the fixed
parameters appear in dashed boxes).}\label{fig:DAG}
\end{figure*}

The posterior distribution of the unknown parameter vector
$\paramvect=\left\{\MATmix,\MATsou,\Vnoisevar\right\}$ can be
computed from the following hierarchical structure:
\begin{equation}
f( \paramvect | \MATobs )  \propto \int f(\MATobs|\paramvect)
f(\paramvect | \hypervect) f(\hypervect) d\hypervect,
\end{equation}
where $f\left(\MATobs\big|\paramvect\right)$ and
$f\left(\hypervect\right)$ have been defined in
\eqref{eq:likelihood} and \eqref{eq:hyper_prior}. Moreover, by
assuming the independence between $\MATmix$, $\MATsou$ and
$\Vnoisevar$, the following result can be obtained:
\begin{equation}
f\left(\paramvect\big|\hypervect\right) =
f\left(\MATmix\right)f\left(\MATsou\big|\Vsouvar\right)f\left(\Vnoisevar\big|\rho_{\mathrm{e}},\psi_{\mathrm{e}}\right),
\end{equation}
where $f\left(\MATmix\right)$, $f\left(\MATsou\big|\Vsouvar\right)$
and
$f\left(\Vnoisevar\big|\rho_{\mathrm{e}},\psi_{\mathrm{e}}\right)$
have been defined previously. This hierarchical structure, depicted
in the directed acyclic graph (DAG) of Fig.~\ref{fig:DAG}, allows
one to integrate out the hyperparameters $\psi_{\mathrm{e}}$ and
$\VsoucoeffB$ from the joint distribution
$f\left(\paramvect,\hypervect|\MATobs\right)$, yielding:
\begin{equation}
\begin{split}
 \label{eq:posterior}
 f&\left(\MATmix,\MATsou,\Vnoisevar,\VsoucoeffA\big|\MATobs\right) \propto
 \prod_{\noobs=1}^\nbobs
    \left[ \frac{\Indicfun{\Simplex}{\Vmix{\noobs}}}{\sigma_{\mathrm{e},i}^{L+2}}
    \exp\left(-\frac{\left\|\Vobs{\noobs}-\MATsou\transp\Vabond{\noobs}\right\|^2}{2\noisevar{\noobs}}\right)\right]\\
 &\times \prod_{\nosou=1}^\nbsou
    \left[\frac{\Gamma\left(\nbband\soucoeffA{\nosou} + \alpha_{\soucoeffB{\nosou}} +1\right)}
        {\left(\sum_{\noband=1}^{\nbband} \sou{\nosou}{\noband} +
            \beta_{\soucoeffB{\nosou}}\right)^{\nbband\soucoeffA{\noband} + \alpha_{\soucoeffB{\nosou}} +1}}\right]\\
 &\times \prod_{\nosou=1}^\nbsou
    \left[\left(\prod_{\noband=1}^{\nbband} \frac{\sou{\nosou}{\noband}}{\Gamma\left(\soucoeffA{\nosou} \right)} \right)^{\soucoeffA{\nosou}-1}
    \Indicfun{\mathbb{R}^\nbband_+}{\Vsou{\nosou}}\right].
\end{split}
\end{equation}
The posterior distribution in \eqref{eq:posterior} is clearly too
complex to derive the classical Bayesian estimators of the unknown
parameters, such as the minimum mean square error (MMSE) estimator
or the maximum \emph{a posteriori} (MAP) estimator. To overcome the
difficulty, it is quite common to make use of MCMC methods to
generate samples asymptotically distributed according to the exact
posterior of interest \cite{Robert1999}. The simulated samples are
then used to approximate integrals by empirical averages for the
MMSE estimator and to estimate the maximum of the posterior
distribution for the MAP estimator. The next section proposes a
Gibbs sampling strategy for the BSS of the spectral mixtures under
the positivity and full additivity constraints.

\section{Gibbs sampler} \label{sec:Gibbs}
The Gibbs sampler is an iterative sampling strategy that consists of
generating samples (denoted $\widetilde{\cdot}^{(t)}$) distributed
according to the conditional distribution of each parameter. This
section describes a Gibbs sampling strategy generating samples
$\left(\widetilde{\MATmix}^{(t)},\widetilde{\MATsou}^{(t)},
\left\{\widetilde{\boldsymbol{\sigma}}^{2}\right\}^{(t)},\widetilde{\boldsymbol{\alpha}}^{(t)}\right)$
asymptotically distributed according to \eqref{eq:posterior}. The
main steps of the algorithm (denoted as Algorithm~\ref{algo:Gibbs})
are detailed from subsection~\ref{subsec:gene_abond} to subsection
\ref{subsec:gene_mat}.
\begin{algogo}{Gibbs sampling algorithm for blind spectral source separation}
   \label{algo:Gibbs}
    \begin{itemize}
    \item \underline{Initialization:}
    \begin{enumerate}
        \item sample the hyperparameter $\samplehypernoise{0}$ from the pdf in \eqref{eq:hypernoise_var_prior},
        \item for $\noobs=1,\ldots,\nbobs$, sample the noise variance
            $\samplenoisevar{\noobs}{0}$ from the pdf in \eqref{eq:var_prior_ind},
        \item for $\nosou=1,\ldots, \nbsou$, sample the
            hyperparameter $\samplesoucoeffA{\nosou}{0}$ from the pdf in
            \eqref{eq:soucoeffA_prior},
        \item for $\nosou=1,\ldots, \nbsou$, sample the
            hyperparameter $\samplesoucoeffB{\nosou}{0}$ from the pdf in
            \eqref{eq:soucoeffB_prior},
        \item for $\nosou=1,\ldots, \nbsou$, sample the source spectrum
            $\sampleVsou{\nosou}{t}$ from the pdf in \eqref{eq:Vsou_prior}.
        \item Set $t \leftarrow 1$,
    \end{enumerate}
    \item \underline{Iterations:} for $t=1,2, \ldots, $ do
    \begin{enumerate}
        \item for $\noobs=1,\ldots,\nbobs$, sample the concentration vector $\sampleVmix{\noobs}{t}$ from the pdf in~\eqref{eq:gene_abond},
        \item sample the hyperparameter $\samplehypernoise{t}$ from the pdf in \eqref{eq:gene_hypernoise},
        \item for $\noobs=1,\ldots,\nbobs$, sample the noise variance
            $\samplenoisevar{\noobs}{t}$ from the pdf in \eqref{eq:gene_noisevar},
        \item for $\nosou=1,\ldots, \nbsou$, sample the hyperparameter
            $\samplesoucoeffA{\nosou}{t}$ from the pdf in \eqref{eq:gene_soucoeffA},
        \item for $\nosou=1,\ldots, \nbsou$, sample the hyperparameter
            $\samplesoucoeffB{\nosou}{t}$ from the pdf in \eqref{eq:gene_soucoeffB},
        \item for $\nosou=1,\ldots, \nbsou$, sample the source spectrum
            $\sampleVsou{\nosou}{t}$ from the pdf in \eqref{eq:gene_mat}.
        \item Set $t \leftarrow t+1$.
    \end{enumerate}
    \end{itemize}
\end{algogo}

\subsection{Generation according to $f\left(\MATmix|\MATsou,\Vnoisevar,\MATobs\right)$}
\label{subsec:gene_abond} Straightforward computations yield for
each observation:
\begin{equation}
f\left(\Vmix{\noobs}\left|\MATsou,\noisevar{\noobs},\Vobs{\noobs}\right.\right)
\propto
    \exp\left[-\frac{\left(\Vmix{\noobs}-\boldsymbol{\mu}_\noobs\right)\transp{\boldsymbol{\Lambda}}^{-1}_\noobs
        \left(\Vmix{\noobs}-\boldsymbol{\mu}_\noobs\right)}{2}\right]\Indicfun{\mathbb{T}}{\Vmix{\noobs}},
        \label{eq:posteriorA}
\end{equation}
where:
\begin{equation}
\label{eq:param_normal} \left\{
\begin{split}
{\boldsymbol{\Lambda}}_\noobs &= \left[\frac{1}{\noisevar{\noobs}}
\left(\MATsou_{\text{--}\nbsou,\boldsymbol{\cdot}}\transp-\Vsou{\nbsou}\un\transp\right)\transp\left(\MATsou_{\text{--}\nbsou,\boldsymbol{\cdot}}\transp-\Vsou{\nbsou}\un\transp\right)\right]^{-1},\\
{\boldsymbol{\mu}}_\noobs& =
 {\boldsymbol{\Lambda}}_\noobs\left[\frac{1}{\noisevar{\noobs}}\left(\MATsou_{\text{--}\nbsou,\boldsymbol{\cdot}}\transp-\Vsou{\nbsou}\un\transp\right)
 \transp\left(\Vobs{\noobs}-\Vsou{\nbsou}\right)\right],
\end{split}
\right.
\end{equation}
with $\un=[1,\ldots,1]\transp \in \mathbb{R}^{\nbsou-1}$ and where
$\MATsou_{\text{--}\nbsou,\boldsymbol{\cdot}}$ denotes the matrix
$\MATsou$ from which the $\nbsou^\mathrm{th}$ row has been removed.
As a consequence,
$\Vmix{\noobs}\big|\MATsou,\noisevar{\noobs},\Vobs{\noobs}$ is
distributed according to a truncated Gaussian distribution on the
simplex $\Simplex$:
\begin{equation}
\label{eq:gene_abond}
\Vmix{\noobs}\left|\MATsou,\noisevar{\noobs},\Vobs{\noobs}\right.
\sim
\mathcal{N}_{\Simplex}\left(\boldsymbol{\mu}_{\noobs},\boldsymbol{\Lambda}_{\noobs}\right).
\end{equation}
When the number $\nbsou$ of spectral sources is relatively small,
the generation of
$\Vmix{\noobs}\left|\MATsou,\noisevar{\noobs},\Vobs{\noobs}\right.$
can be achieved using a standard Metropolis Hastings (MH) step. By
choosing the Gaussian distribution
$\mathcal{N}\left(\boldsymbol{\mu}_{\noobs},\boldsymbol{\Lambda}_{\noobs}\right)$
as proposal distribution for this MH step, the acceptance ratio of
the MH algorithm reduces to $1$ if the candidate is inside the
simplex $\Simplex$ and $0$ otherwise. For higher dimension problems,
the acceptance ratio of the MH algorithm can be small, leading to
poor mixing properties. In such cases, an alternative strategy based
on a Gibbs sampler can be used (see \cite{Robert1995} and
\cite{DobigeonTR2007b}).

\subsection{Generation according to $f\left(\Vnoisevar\left|\MATmix,\MATsou,\MATobs\right.\right)$}
To sample according to
$f\left(\Vnoisevar\left|\MATmix,\MATsou,\MATobs\right.\right)$, it
is very convenient to generate samples from
$f\left(\Vnoisevar,\psi_{\mathrm{e}}\left|\MATmix,\MATsou,\MATobs\right.\right)$
by using the two following steps:

\subsubsection{Generation according to $f\left(\psi_{\mathrm{e}}\left|\Vnoisevar,\MATmix,\MATsou,\MATobs\right.\right)$}
The conditional distribution is expressed as the following IG
distribution:
\begin{equation}
\label{eq:gene_hypernoise}
  \psi_\mathrm{e}\left|\Vnoisevar,\rho_\mathrm{e}\right. \sim
  \mathcal{IG}\left(\frac{\nbobs\rho_\mathrm{e}}{2},\frac{1}{2}\sum_{\noobs=1}^{\nbobs}\frac{1}{\noisevar{\noobs}}\right).
\end{equation}

\subsubsection{Generation according to $f\left(\Vnoisevar\left|\psi_{\mathrm{e}},\MATmix,\MATsou,\MATobs\right.\right)$}
After a careful examination of
$f\left(\Vnoisevar,\MATmix,\psi_{\mathrm{e}}\big
|\MATsou,\MATobs\right)$, it can be deduced that the conditional
distribution of the noise variance in each observation spectrum is
the following IG distribution:
\begin{equation} \label{eq:gene_noisevar}
\noisevar{\noobs}\left|\psi_{\mathrm{e}},\Vmix{\noobs},\MATsou,\Vobs{\noobs}\right.
\sim \mathcal{IG}\left(\frac{\rho_{\mathrm{e}}+\nbband}{2},
 \frac{\psi_{\mathrm{e}}+\left\|\Vobs{\noobs}-\MATsou\Vabond{\noobs}\transp\right\|^2}{2}\right).
\end{equation}

\subsection{Generation according to $f\left(\MATsou\left|\MATmix,\Vnoisevar,\MATobs\right.\right)$}
\label{subsec:gene_mat} This generation can be achieved thanks to
the three following steps, as in \cite{Moussaoui2006b}.

\subsubsection{Generation according to $f\left(\VsoucoeffA
    \left|\VsoucoeffB,\MATsou,\MATmix,\Vnoisevar,\MATobs\right.\right)$}
From the joint distribution
$f\left(\MATmix,\MATsou,\Vnoisevar,\VsoucoeffA,\VsoucoeffB|\MATobs\right)$,
we can express the posterior distribution of $\soucoeffA{\nosou}$
($\nosou =1,\ldots,\nbsou$) as:
\begin{equation}
\label{eq:gene_soucoeffA}
  f\left(\soucoeffA{\nosou} \big|\Vsou{\nosou}, \soucoeffB{\nosou}\right) \propto
    \prod_{\noband=1}^{\nbband} \left[\frac{\soucoeffB{\nosou}^{\soucoeffA{\nosou}}}{\Gamma\left(\soucoeffA{\nosou}\right)}
        \sou{\nosou}{\noband}^{\soucoeffA{\nosou}}\right]
        e^{-\lambda_{\soucoeffA{\nosou}}\soucoeffA{\nosou}}\Indicfun{\mathbb{R}^+}{\soucoeffA{\nosou}}.
\end{equation}

This posterior is not easy to simulate as it does not belong to a
known distribution family. Therefore, an MH step is required to
generate samples $\widetilde{\alpha}_{\nosou}^{(t)}$ distributed
according to \eqref{eq:gene_soucoeffA}. The reader is invited to
consult \cite{Moussaoui2006b} for more details regarding the choice
of the instrumental distribution in order to obtain a high
acceptance rate for the MH algorithm.

\subsubsection{Generation according to $f\left(\VsoucoeffB
    \left|\VsoucoeffA,\MATsou,\MATmix,\Vnoisevar,\MATobs\right.\right)$}
Similarly, the posterior distribution of the hyperparameter vector
$\VsoucoeffB$ can be determined by looking at the joint distribution
$f\left(\MATmix,\MATsou,\Vnoisevar,\VsoucoeffA,\VsoucoeffB|\MATobs\right)$.
In this case, the posterior distribution of the individual
hyperparameter $\soucoeffB{\nosou}$ ($\nosou =1,\ldots,\nbsou$) is
the following Gamma distribution:
\begin{equation}
\label{eq:gene_soucoeffB}
  \soucoeffB{\nosou} \big|  \soucoeffA{\nosou}, \Vsou{\nosou} \sim
  \mathcal{G}\left(1 + \nbband  \soucoeffA{\nosou} + \alpha_{\soucoeffA{\nosou}},\sum_{\noband=1}^{\nbband} \sou{\nosou}{\noband} +
    \beta_{\soucoeffA{\nosou}}\right).
\end{equation}

\subsubsection{Generation according to $f\left(\MATsou\left|\VsoucoeffA,\VsoucoeffB,\MATmix,\Vnoisevar,\MATobs\right.\right)$}
Finally, the posterior distribution of the source observed in the
$\noband^{\text{th}}$ spectral band is:
\begin{equation}\label{eq:gene_mat}
  f\left(\sou{\nosou}{\noband} \big|\soucoeffA{\nosou},\soucoeffB{\nosou},\MATmix,\Vnoisevar,\MATobs \right)
    \propto \sou{\nosou}{\noband}^{\soucoeffA{\nosou}-1}
    \Indicfun{\mathbb{R}^+}{\sou{\nosou}{\noband}}
    \exp\left[-\frac{\left(\sou{\nosou}{\noband}-\mu_{\nosou,\noband}\right)^2}{2\delta^2_{\nosou}} - \soucoeffB{\nosou} \sou{\nosou}{\noband}\right],
\end{equation}
with
\begin{equation}
    \label{eq:param_post_sou}
    \left\{
    \begin{array}{l}
    \delta^2_{\nosou}=\left[\sum_{\noobs=1}^{\nbobs}
        \frac{\abond{\noobs}{\nosou}^2}{\noisevar{\noobs}}\right]\inv,\\
    \mu_{\nosou,\noband}=\frac{1}{\delta^2_{\nosou}}
        \sum_{\noobs=1}^{\nbobs} \frac{\abond{\noobs}{\nosou}\epsilon_{\noobs,\noband}^{(-\nosou)}}{\noisevar{\noobs}},
    \end{array}\right.
\end{equation}
where $\epsilon_{\noobs,\noband}^{(-\nosou)}=\obs{\noobs}{\noband} -
\sum_{k \neq \nosou} \abond{\noobs}{k}
 \sou{k}{\noband}$.
The generation of samples distributed according to
\eqref{eq:gene_mat} is achieved by using an MH algorithm whose
proposal is a positive truncated normal distribution
\cite{Moussaoui2006b}. The generation according to the positive
truncated Gaussian distribution can be achieved thanks to an
accept-reject scheme with multiple proposal distributions (see
\cite{Geweke1991,Robert1995,Mazet2005} for details).

\section{Experimental results with synthetic data} \label{sec:simus_synth}
This section presents some experiments performed on synthetic data
to illustrate the performance of the proposed Bayesian spectral
unmixing algorithm.

\subsection{Mixture synthesis}
\begin{figure*}[t!]
\centering{
\includegraphics[width=\textwidth]{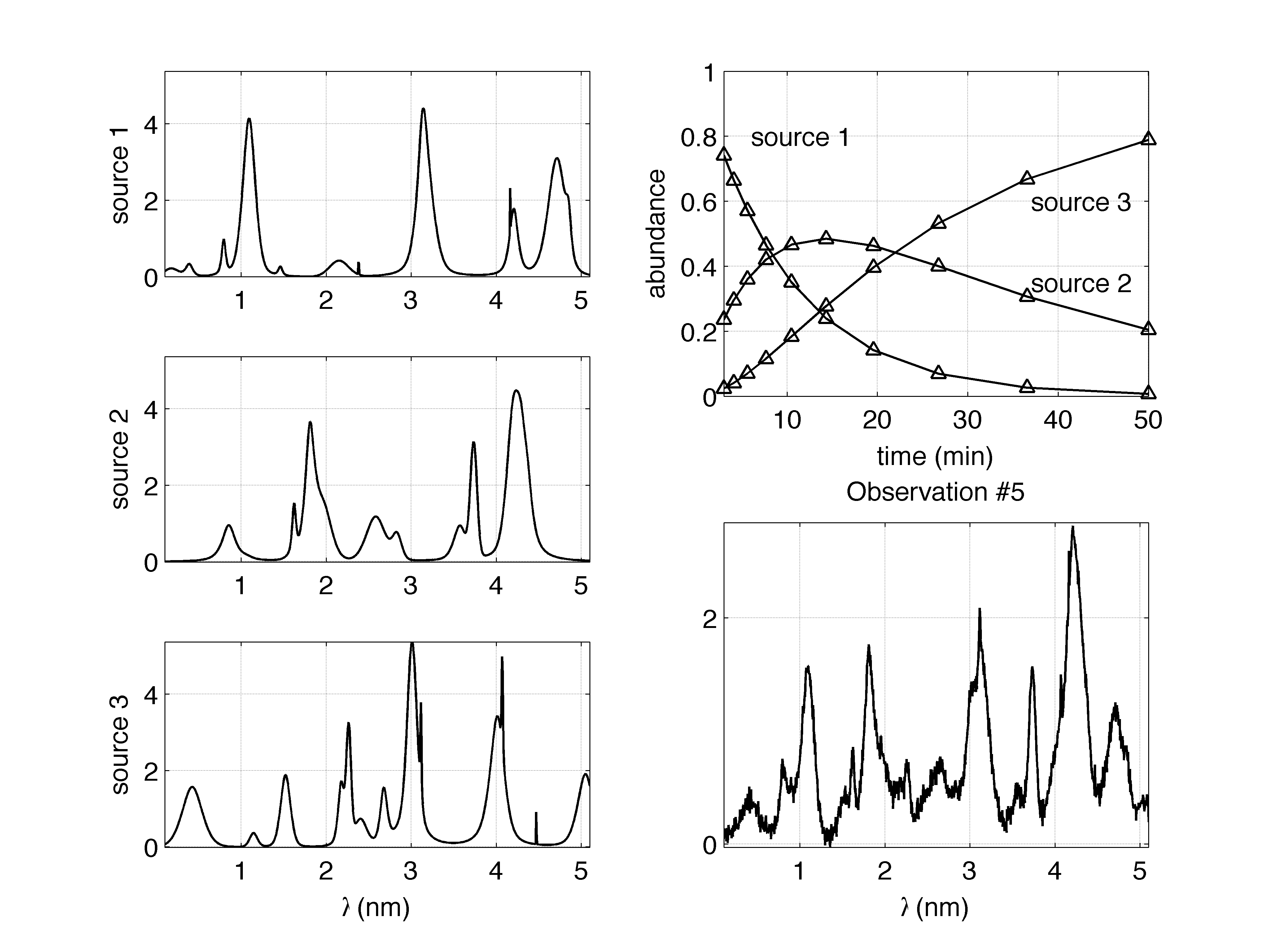}
\caption{Left: example of $\nbsou=3$ simulated spectral sources
where the $x$-axis corresponds to the wavelength expressed in $nm$
and the $y$-axis corresponds to the absorbance of the spectra.
Right, top: abundance evolution profiles. Right, bottom: one typical
realization of the observed spectra. \label{fig:example}}}
\end{figure*}

The spectral sources have been simulated to get signals similar to
absorption spectroscopy data. Each spectrum is obtained as a
superposition of Gaussian and Lorentzian functionals with randomly
chosen parameters (location, amplitude and width)
\cite{Moussaoui2006b}. Figure \ref{fig:example} (left) shows an
example of  $\nbsou = 3$ source signals of $\nbband = 1000$ spectral
bands. For this application, a ``spectral" band corresponds to a
given value of the wavelength $\lambda$ (expressed in nanometers).
The mixing coefficients have been chosen to obtain evolution
profiles similar to component abundance variation in a kinetic
reaction, as depicted in Figure \ref{fig:example} (top, right). The
abundance fraction profiles have been simulated for $\nbobs=10$
observation times, which provides $\nbobs=10$ observation spectra.
An i.i.d. Gaussian sequence has been added to each observation with
appropriate standard deviation to have a signal to noise ratio (SNR)
equal to $20$dB. One typical realization of the observed spectra is
shown in Figure \ref{fig:example} (bottom, right).

\subsection{Separation with non-negativity and full additivity constraints}
\begin{figure*}[t!]
\centering{
\includegraphics[width=\textwidth]{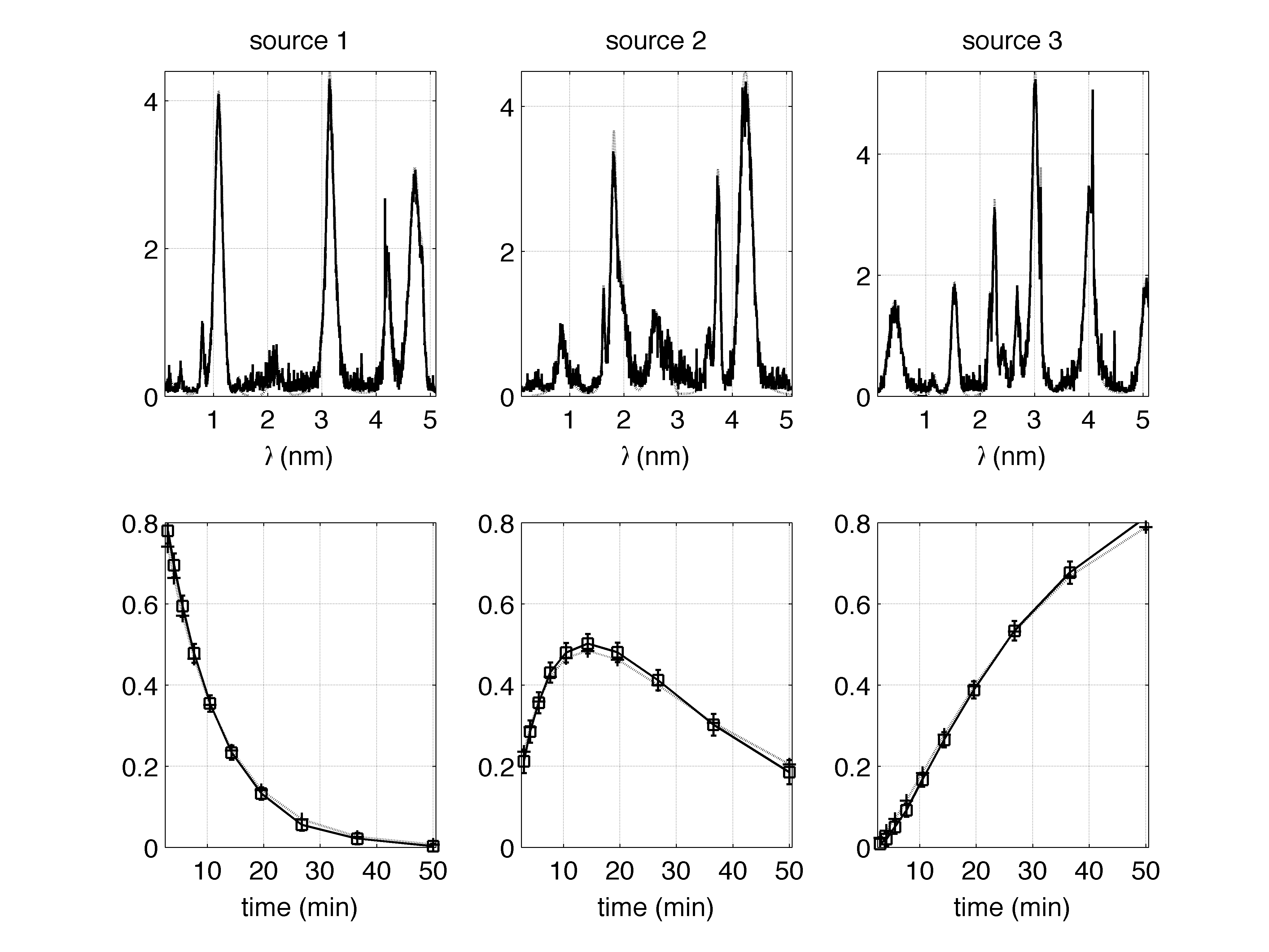}
\caption{Top: Simulated (dotted) and estimated (continuous line)
source spectra. Bottom: Simulated values (cross) and MMSE estimates
(circles) of the abundances. Error bars indicate the estimated 95\%
confidence intervals from the simulated Markov chain. }
\label{fig:resultat_gamma}}
\end{figure*}

Figure~\ref{fig:resultat_gamma} summarizes the result of a Monte
Carlo simulation with $100$ runs where the mixing matrix has been
kept unchanged, while new sources and noise sequences have been
generated at each run. Figure~\ref{fig:resultat_gamma}-a shows a
comparison between the true concentrations (cross) and their MMSE
estimates (circles) obtained for a Markov chain of $N_{\textrm{MC}}
= 1000$ iterations including $N_{\textrm{b-i}}=200$ burn-in
iterations. These estimates have been computed according to the MMSE
principle ($\noobs = 1,\ldots,\nbsou$):
\begin{equation}\label{eq:abond_MMSE}
 \hat{\Vmix{}}_\noobs = \frac{1}{N_r} \sum_{t=1}^{N_r}
 \sampleVmix{\noobs}{N_{\textrm{b-i}}+t},
\end{equation}
where $N_r = N_{\textrm{MC}} - N_{\textrm{b-i}}$ is the number of
iterations used for the estimation. The estimated abundances are
clearly in good agreement with the actual abundances and the
estimates satisfy the positivity and full additivity constraints. By
comparing figures \ref{fig:example} (left) and
\ref{fig:resultat_gamma} (top), it can be observed that the source
signals have also been correctly estimated.

\begin{figure}[h!]
\centering{
\includegraphics[width=\halffigwidth]{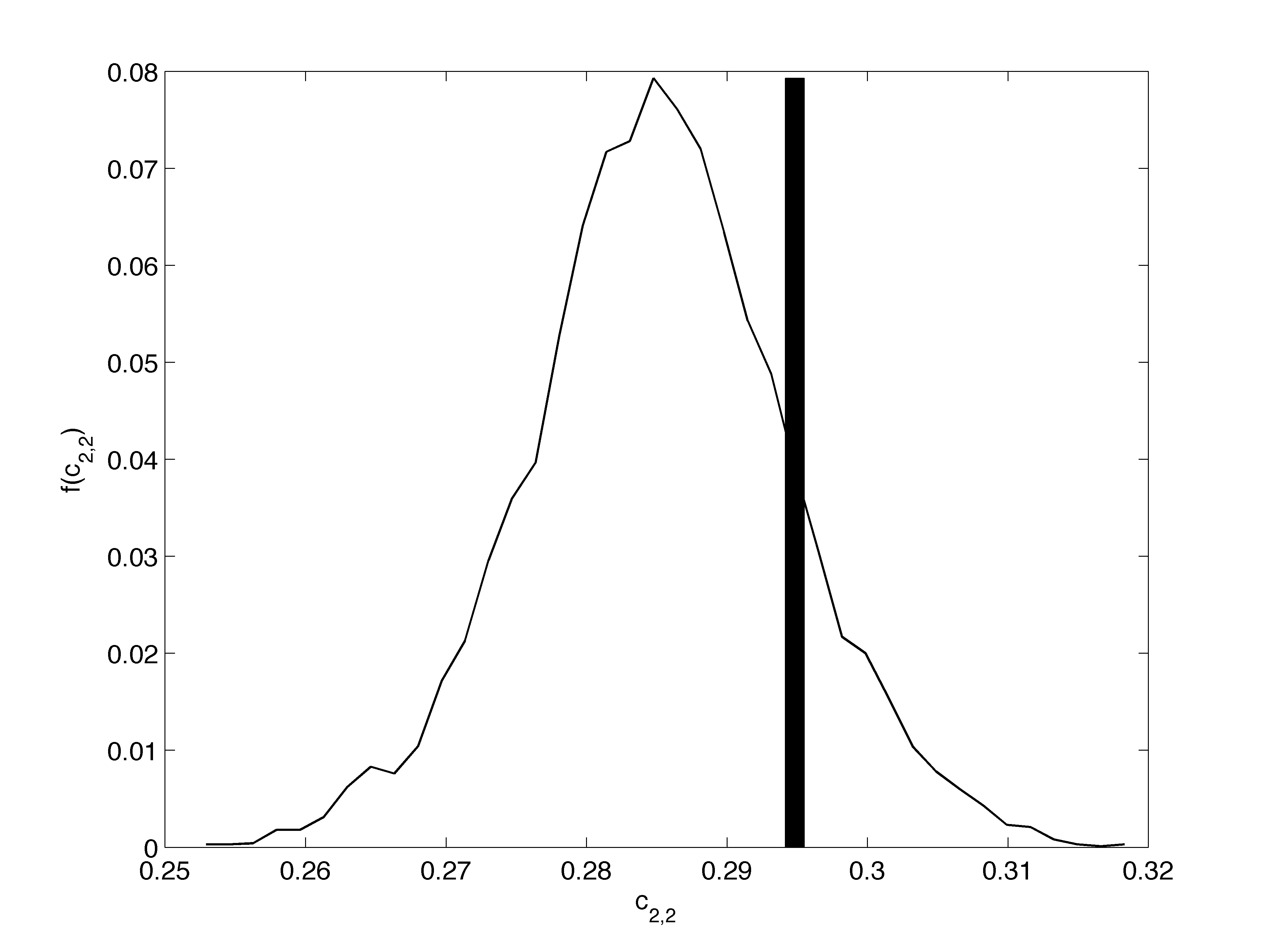}
\includegraphics[width=\halffigwidth]{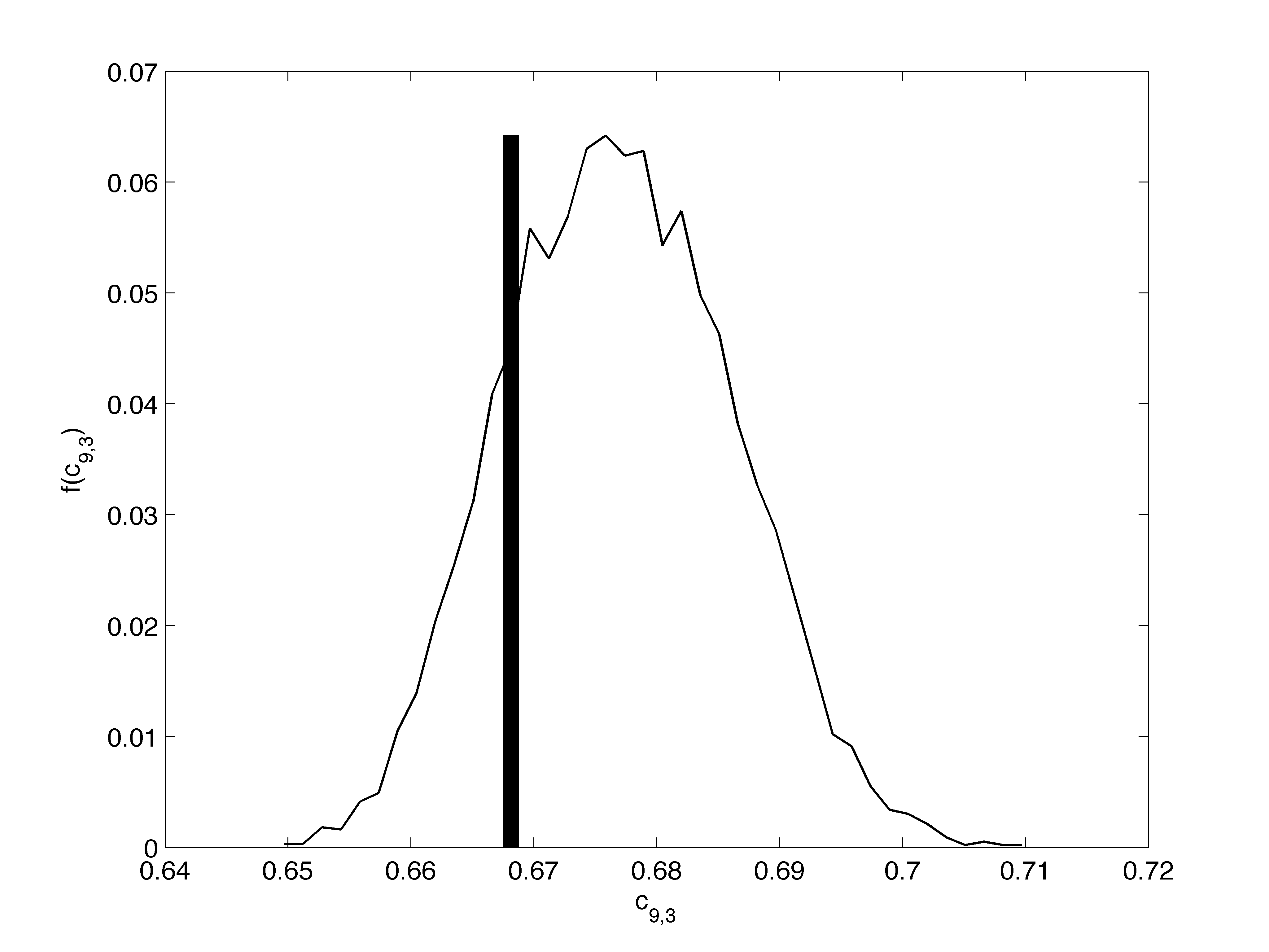}
\caption{Left (resp. right): posterior distribution of the
concentration of the $2^\text{nd}$ (resp. $3^\text{rd}$) spectral
component in the mixture observed at index time $\noobs=2$ (resp.
$\noobs=9$). The actual values appear as black bars.}
\label{fig:resultat_conc_post}}
\end{figure}

It is interesting to note that the proposed algorithm generates
samples distributed according to the posterior distribution of the
unknown parameters
$f\left(\MATmix,\MATsou,\Vnoisevar,{\boldsymbol{\alpha}}\big|\MATobs\right)$.
These samples can be used to obtain the posterior distributions of
the concentrations or the source spectra. As an example, typical
posterior distributions for two mixing coefficients are depicted in
figure~\ref{fig:resultat_conc_post}. These posteriors are in good
agreement with the theoretical posterior distributions in
\eqref{eq:gene_abond}, i.e. truncated Gaussian distributions.

\subsection{Monitoring sampler convergence}
An important issue when using MCMC methods is convergence
monitoring. The Gibbs sampler detailed in section \ref{sec:Gibbs}
generates random samples
$\left(\widetilde{\MATmix}^{(t)},\widetilde{\MATsou}^{(t)},
\widetilde{\boldsymbol{\sigma}}^{2(t)},\widetilde{\boldsymbol{\alpha}}^{(t)}\right)$
asymptotically distributed according to the posterior distribution
in \eqref{eq:posterior}. The quantities of interest, i.e. the
concentration coefficients and the source spectra, are then
approximated by empirical averages according to
\eqref{eq:abond_MMSE}. However, two essential parameters have to be
tuned: the length $N_{\mathrm{MC}}$ of the constructed Markov chain
and the length $N_{\mathrm{b-i}}$ of the burn-in period, i.e. the
number of simulated samples to be discarded before computing the
averages. This section reports some works conducted to ensure the
convergence of the proposed algorithm and the accuracy of the
estimation for the unknown parameters.

First, the burn-in period $N_\textrm{b-i}=200$ has been determined
thanks to the popular potential scale reduction factor (PSRF). The
PSRF was introduced by Gelman and Rubin \cite{Gelman1992} and has
been widely used in the signal processing literature (see for
instance \cite{Djuric2002,Dobigeon2007a,Dobigeon2007b}). It consists
of running several parallel Markov chains and computing the
following criterion:
\begin{equation}
  \hat \rho =
 \left (1- \frac{1}{N_r}\right) \left[1+ \frac{1}{(N_r-1)} \frac{B(\kappa)}{W(\kappa)}\right],
\end{equation}
where $W$ and $B$ are the within and between-sequence variances of
the parameter $\kappa$, respectively. Different choices for $\kappa$
can be used for our source separation problem. Here, we consider the
parameters $\noisevar{\noobs}$ ($\noobs = 1,\ldots,\nbobs$) as
recommended in \cite{Godsill1998}. Table~\ref{tab:convergence} shows
the PSRF obtained for the $\nbobs=10$ observation times computed
from $M=10$ Markov chains. All these values of $\sqrt{\hat \rho}$
confirm the good convergence of the sampler since a recommendation
for convergence assessment is $\sqrt{\hat \rho}<1.2$ \cite[p.
332]{Gelman1995}.


\begin{table}[h!]
\renewcommand{\arraystretch}{1.8}
\centering \caption{Potential scale reduction factors of
$\noisevar{\noobs}$ computed from $M=10$ Markov chains.}
\begin{tabular}{|cc||cc|}
 \hline
 Obs. index & $\sqrt{\hat \rho}$ & Obs. index & $\sqrt{\hat \rho}$ \\
 \hline
 1 & 1.0048 & 2 & 1.0013 \\
 3 & 1.0027 & 4 & 0.9995 \\
 5 & 1.0097 & 6 & 1.0078 \\
 7 & 1.0001 & 8 & 0.9994 \\
 9 & 1.0080 & 10 & 1.0288\\
 \hline
\end{tabular}
 \label{tab:convergence}
\end{table}

\begin{figure}[h!]
\centering{
\includegraphics[width=0.5\textwidth]{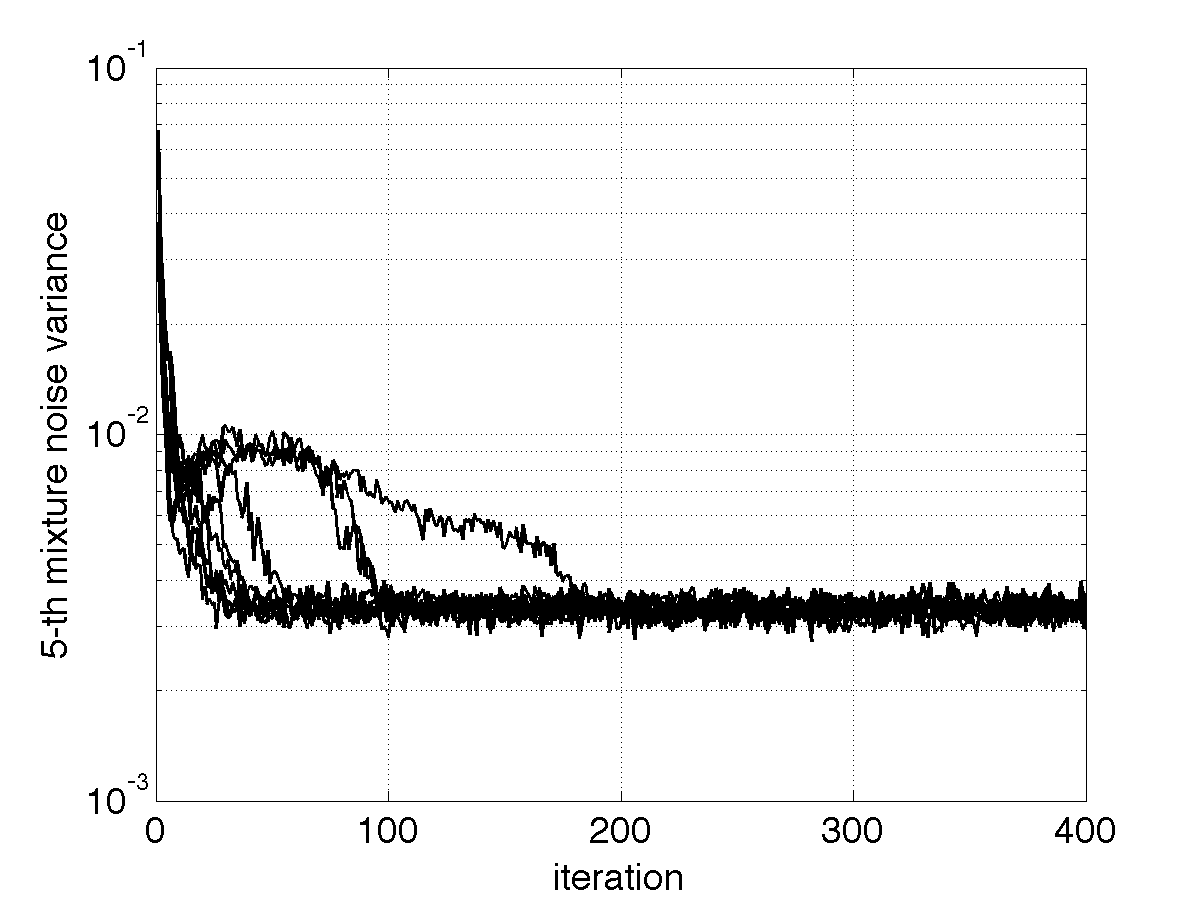}
\caption{Outputs of $M = 10$ Markov chains for the parameter
$\noisevar{5}$.} \label{fig:sigma_output}}
\end{figure}

The Markov chain convergence can also be monitored by a graphical
supervision of the generated samples of the noise variances. As an
illustration, the outputs of $10$ Markov chains for one of the
parameter $\noisevar{\noobs}$ are depicted in
figure~\ref{fig:sigma_output}. All the generated samples converge to
a similar value after a short burn-in period ($200$ iterations, in
this example).

\begin{figure}[h!]
\centering{
\includegraphics[width=0.5\textwidth]{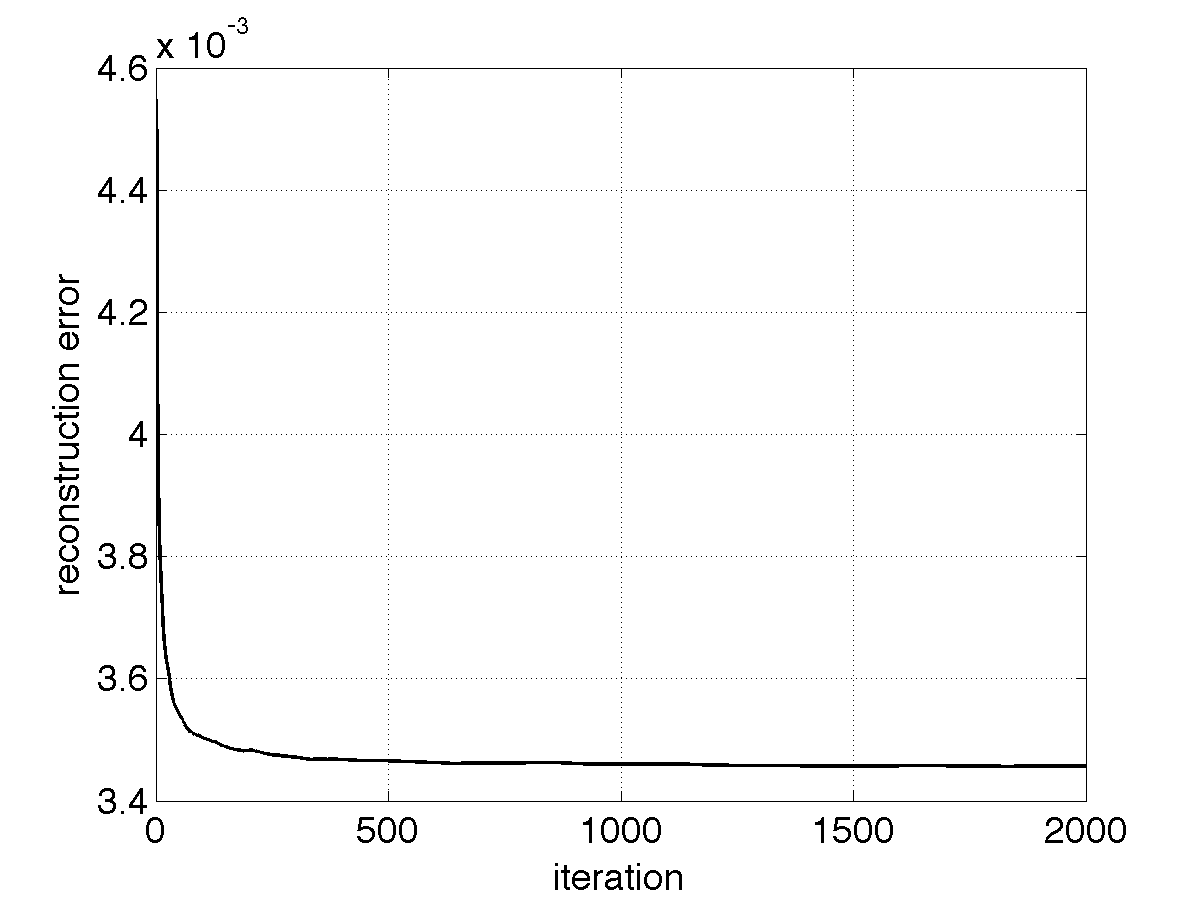}
\caption{Evolution of the reconstruction error with respect to the
iteration number (with a burn-in of $N_{\textrm{b-i}}=200$
iterations).} \label{fig:erreur}}
\end{figure}

Once the number of burn-in iterations has been fixed, the number of
iterations necessary to obtain accurate estimates of the unknown
parameters via \eqref{eq:abond_MMSE} has to be adjusted. This paper
proposes to evaluate $N_r$ with appropriate graphical evaluations
(see \cite[p. 28]{Robert1998} for motivations).
Figure~\ref{fig:erreur} shows the reconstruction error associated to
the different spectra defined as:
\begin{equation}
  e_r^2(p) = \dfrac{1}{N L} \sum_{\noobs=1}^{\nbobs} \left\|\Vobs{\noobs} -
  \left(\hat{\Vabond{\noobs}}^{(p)}\hat{\MATsou}^{(p)}\right)\transp\right\|^2,
\end{equation}
where $\hat{\Vabond{\noobs}}^{(p)}$ and $\hat{\MATsou}^{(p)}$ are
the MMSE estimates of the abundance vector $\Vabond{\noobs}$ and the
source matrix $\MATsou$ computed after $N_{\textrm{b-i}}=200$
burn-in iterations and $N_r=p$ iterations. The number of iterations
$N_r$ required to compute the empirical averages following the MMSE
estimator \eqref{eq:abond_MMSE} can be fixed to ensure the
reconstruction error is below a predefined threshold.
Figure~\ref{fig:erreur} shows that a number of iterations $N_r=500$
is sufficient to ensure a good estimation of the quantities of
interest $\MATabond$ and $\MATsou$.

\subsection{Comparison with other BSS algorithms} \label{subsec:comparisons}
The proposed Bayesian approach has been compared with other standard
BSS methods. Synthetic mixtures have been processed by the
non-negative ICA (NN-ICA) algorithm proposed by Plumbley and Oja
\cite{Plumbley2004}, the iterative NMF method described in
\cite{Sajda2004} and  the Bayesian Positive Source Separation (BPSS)
algorithm introduced in \cite{Moussaoui2006b}.

All these methods do not include the full additivity constraint. To
evaluate the relevance of this additional constraint, \emph{ad hoc}
re-scaled versions of these methods have also been considered.
Simulations have been conducted by applying the $4$ algorithms using
$100$ Monte Carlo runs, each run being associated to a randomly
generated source. Table \ref{tab:comparison} shows the normalized
mean square errors (NMSEs) for the estimated sources and abundance
matrices as defined in \cite{Tugnait1997}:

\begin{equation}
\begin{split}
    \label{eq:NMSE}
    \mathrm{NMSE}\left(\MATsou\right) &= \sum_{\nosou=1}^{\nbsou} \frac{\left\|\Vsou{\nosou} -
    \hat{\mathbf{s}}_{\nosou}\right\|^2}{
    \left\|\Vsou{\nosou}\right\|^2},\\
    \mathrm{NMSE}\left( \MATabond \right) &= \sum_{\noobs=1}^{\nbobs}\frac{ \left\|\Vabond{\noobs} -
    \hat{\mathbf{c}}_{\noobs}\right\|^2}{
    \left\|\Vabond{\noobs}\right\|^2}.
\end{split}
\end{equation}

In addition, the estimation performances have been compared in terms
of dissimilarity. Denoted $\mathrm{diss}\left(\cdot,\cdot\right)$,
it measures how the estimated source spectrum differs from the
reference one \cite{Moussaoui2006a} and is defined by:
\begin{equation}
    \mathrm{diss}\left(\Vsou{m},\hat{\bfs}_{m}\right) = \sqrt{1 - \mathrm{corr}\left(\Vsou{m},\hat{\bfs}_{m}\right)^2},
\end{equation}
where  $\mathrm{corr}(\bfs_m,\hat \bfs_m)$ is the correlation coefficient between $\bfs_m$ and its estimate $\hat \bfs_m$. Consequently the average dissimilarity over the $\nbsou$ sources is
reported in Table \ref{tab:comparison}.

\begin{table}[h!]
\renewcommand{\arraystretch}{1.8}
\centering \caption{Estimation performance for different BSS
algorithms ($100$ Monte Carlo runs). }
\begin{tabular}{lcccc}
 \cline{1-5} \multicolumn{1}{c}{} &  $\mathrm{NMSE}\left(\MATsou\right)$ & $\mathrm{NMSE}\left(\MATabond\right)$  & Av. $\mathrm{Diss}\left(\MATsou\right)$ & Time (min)\\
 \hline
 Proposed approach &  0.0071 & 0.0024 & 13.9 \%   & 44\\
 BPSS               & 0.0121 & 0.0025 & 13.4 \%   & 45\\
 re-scaled BPSS     & 0.0126 & 0.0023 & 13.4 \%   & 45\\
 NN-ICA             & 0.0613 & 0.0345 & 20.0 \%   & 3\\
 re-scaled NN-ICA   & 0.0602 & 0.0384 & 19.4 \%   & 3\\
 NMF                & 0.2109 & 1.9149 & 22.6 \%   & 1\\
 re-scaled NMF      & 0.0575 & 0.0496 & 24.5 \%   & 1\\
 \hline
\end{tabular}
 \label{tab:comparison}
\end{table}

These results demonstrate that an \emph{ad hoc} re-scaling of the
results obtained by NMF techniques is not always an efficient means
to improve the estimation performance. Indeed, the \emph{ad hoc}
re-scaled version of NMF provides lower MSEs than the corresponding
standard algorithms. On the other hand, this constraint does not
significantly improve the NN-ICA or the BPSS algorithms. As far as
the Bayesian algorithms are concerned, they clearly provide better
estimation performance than the non-Bayesian approaches. However,
the proposed fully constrained algorithm clearly outperforms the two
BPSS algorithms, especially regarding the source estimation.

The computation times required by each of these algorithms are also
reported in Table~\ref{tab:comparison} for a MATLAB implementation
on a $2.2$GHz Intel Core 2. This shows that the complexities of the
proposed method and BPSS algorithms are quite similar and higher
than the complexities of the NN-ICA and MNF algorithms. This seems
to be the price to pay to obtain significantly better estimation
performances.


\subsection{Modified Bayesian models with other source priors}
\label{subsec:modified_BM} As it has been mentioned previously,
several distributions can be chosen as priors for the source
spectra, provided these distributions have positive supports. The
previous HBM studied in section \ref{subsubsec:sources_prior} is
based on Gamma distributions as source priors. However, simpler
models can be obtained for instance by choosing exponential priors
with different scale parameters $\souvar{\nosou}$:
\begin{equation}
  f\left(\Vsou{\nosou}\big|\souvar{\nosou}\right) \propto
  \prod_{\noband=1}^{\nbband} \frac{1}{\souvar{\nosou}}
  \exp\left(-\frac{\sou{\nosou}{\noband}}{2\souvar{\nosou}}
  \right) \Indicfun{\mathbb{R}^+}{\sou{\nosou}{\noband}},
\end{equation}
or positive truncated Gaussian distribution with different hidden variances $\souvar{\nosou}$:
\begin{equation}
  f\left(\Vsou{\nosou}\big|\souvar{\nosou}\right) \propto
  \prod_{\noband=1}^{\nbband}\frac{1}{\souvar{\nosou}}
  \exp\left(-\frac{\sou{\nosou}{\noband}^2}{2\souvar{\nosou}} \right)
  \Indicfun{\mathbb{R}^+}{\sou{\nosou}{\noband}}.
\end{equation}
The resulting Bayesian algorithms are simpler since only one hyperparameter $\souvar{\nosou}$ has to be adjusted for each source.

For both choices, conjugate IG distributions
$\mathcal{IG}\left(\frac{\rho_{\mathrm{s}}}{2},\frac{\psi_{\mathrm{s}}}{2}\right)$
are chosen as prior distributions for the hyperparameters
$\souvar{\nosou}$, $\nosou=1,\ldots,\nbsou$. After integrating out
the hyperparameter vector
$\hypervect=\left\{\psi_{\mathrm{e}},\Vsouvar\right\}$, the
posterior distribution in \eqref{eq:posterior} can be expressed as:
\begin{equation}
\label{eq:posterior_exponential}
 f\left(\MATmix,\MATsou,\Vnoisevar\big|\MATobs\right) \propto
  \prod_{\nosou=1}^{\nbsou}
  T \left(\Vsou{\nosou},\rho_{\mathrm{s}},\psi_{\mathrm{s}}\right)
    \Indicfun{\mathbb{R}^\nbband_+}{\Vsou{\nosou}}
    \prod_{\noobs=1}^{\nbobs} \left[ \frac{1}{\sigma_{e,i}^{L+2}}
    \exp \left(-\frac{\left\|\Vobs{\noobs}-\MATsou\transp\Vabond{\noobs}\right\|^2}{2\noisevar{\noobs}}\right)\Indicfun{\Simplex}{\Vmix{\noobs}} \right].
\end{equation}
The scalar
$T\left(\Vsou{\nosou},\rho_{\mathrm{s}},\psi_{\mathrm{s}}\right)$
depends on the prior distribution used for the source spectra:
\begin{equation}
T\left(\Vsou{\nosou},\rho_{\mathrm{s}},\psi_{\mathrm{s}}\right) =
\left\{
\begin{array}{ll}
     \left[\psi_{\mathrm{s}}+
   \left\|\Vsou{\nosou}\right\|_1\right]^{-\frac{\nbband+\rho_{\mathrm{s}}}{2}},
        & \hbox{ for exponential priors,} \\
     \left[\psi_{\mathrm{s}}+
   \left\|\Vsou{\nosou}\right\|_2^2\right]^{-\frac{\nbband+\rho_{\mathrm{s}}}{2}},
        & \hbox{ for truncated Gaussian priors.} \\
\end{array}
\right.
\end{equation}

In the Gibbs sampling strategy presented in section~\ref{sec:Gibbs},
the generation according to
$f\left(\MATsou\left|\MATmix,\Vnoisevar,\MATobs\right.\right)$ in
subsection~\ref{subsec:gene_mat} is finally achieved using the
following two steps:
\begin{itemize}
  \item generation according to
  $f\left(\Vsouvar\left|\MATsou,\MATmix,\Vnoisevar,\MATobs\right.\right)$:
        \begin{equation}
        \label{eq:gene_souvar_exponential}
        \souvar{\nosou}\big|\Vsou{\nosou} \sim
        \mathcal{IG}\left(\nbband+\rho_{\mathrm{s}},\psi_{\mathrm{s}}+\left\|\Vsou{\nosou}\right\|^b_{\ell_b}\right),
        \end{equation}
  where $b = 1$ for the exponential prior and $b =2$ otherwise,
  \item generation according to $f\left(\MATsou\left|\Vsouvar,\MATmix,\Vnoisevar,\MATobs\right.\right)$
    \begin{equation}
        \label{eq:gene_mat_exponential}
    \Vsou{\nosou}\left|\Vsouvar,\MATmix,\Vnoisevar,\MATobs\right.
    \sim
    \mathcal{N}^{+}\left(\boldsymbol{\lambda}_\nosou,\delta_\nosou^2\Id_{\nbband}\right),
    \end{equation}
  where $\boldsymbol{\lambda}_\nosou$ and $\delta_\nosou^2$, similar to \eqref{eq:param_post_sou},
  are derived following the model in \cite{Moussaoui2006b}.
\end{itemize}

Table \ref{tab:montecarlo} reports the NMSEs (computed from $100$
Monte Carlo runs following \eqref{eq:NMSE}) for the sources and
concentration matrices estimated by the different Bayesian
algorithms. The results are significantly better when employing the
Gamma distribution, which clearly indicates that the Gamma prior
seems to be the best choice to model the distribution of the sources
when analyzing spectroscopy data.

\begin{table}[h!]
\renewcommand{\arraystretch}{1.8}
\centering \caption{NMSE for different source priors ($100$ Monte
Carlo runs). }
\begin{tabular}{cccc}
 \cline{1-4} \multicolumn{1}{c}{} & Gamma & Truncated Gaussian & Exponential \\
 \hline
  $\mathrm{NMSE}\left(\MATsou\right)$ &  0.0071 & 0.0269 & 0.0110 \\
 $\mathrm{NMSE}\left(\MATabond\right)$ & 0.0024 &   0.0089  &   0.0029\\
 \hline
\end{tabular}
 \label{tab:montecarlo}
\end{table}

\section{Separation of chemical mixtures monitored by Raman spectroscopy} \label{sec:simus_real}

Calcium carbonate is a chemical material used commercially for a
large variety of applications such as filler for plastics or paper.
Depending on operating conditions, calcium carbonate crystallizes as
calcite, aragonite or vaterite. Calcite is the most
thermodynamically stable of the three, followed by aragonite or
vaterite. Globally, the formation of calcium carbonate by mixing two
solutions containing respectively calcium and carbonate ions takes
place in two well distinguished steps. The first step is the
precipitation one. This step is very fast and provides a mixture of
calcium carbonate polymorphs\footnote{The ability of a chemical
substance to crystallize with several types of structures, depending
on a physical parameter, such as temperature, is known as
polymorphism. Each particular form is said a polymorph.}. The second
step (a slow process) represents the phase transformation from the
unstable polymorphs to the stable one (calcite). The physical
properties of the crystallized product depend largely on the
polymorphic composition, so it is necessary to quantify these
polymorphs when they are mixed. Several techniques based on infrared
spectroscopy (IR), X ray-diffraction (XRD) or Raman spectroscopy
(RS) can be used to determine the composition of CaCO$_3$ polymorph
mixtures. However, contrary to XRD and IR, RS is a faster method
since it does not require a sample preparation and is a promising
tool for an online polymorphic composition monitoring. In our case,
the crystallization process of calcium carbonate is carried out in
$5\textrm{mol$/$L}$ NaCl solutions, which correspond to a real
industrial situation. Under the industrial conditions, the calcite
is the desired product.

The main purpose of this experiment is to show how the proposed
constrained BSS method can be used for processing Raman spectroscopy
data to study the relation between polymorphs and temperature and to
explore favorable conditions for calcite formation in saline
solutions.

\subsection{Mixture preparation and data acquisition}
Calcium chloride and sodium carbonate separately dissolved in sodium
chloride solutions of the same concentration ($5\textrm{mol$/$L}$)
were rapidly mixed to precipitate calcium carbonate. A $100$mL
solution containing $0.625$M of Na$_2$CO$_3$ and $5$M of NaCl was
added to a $2.5$L solution containing $0.025$M of CaCl$_2$ and $5$M
of NaCl (the precipitation is carried out under stoichiometric
conditions). A preliminary investigation detailed in
\cite{Carteret2009} suggested that the temperature and the aging
time are the most important factors that can affect the polymorphic
composition. Therefore the experiments were operated in a
temperature range between $20$°C and $70$°C and retaining several
aging times of the precipitated mixture. A sample was collected $2$
minutes after the beginning of the experiment to determine the
polymorphic composition at the end of the precipitation step. Then,
samples were collected at regular time intervals to follow the
polymorph transformation.

Raman Spectra were collected on a Jobin-Yvon T$64000$ spectrometer
equipped with an optical microscope, a threefold monochromator, and
a nitrogen-cooled CCD camera. The excitation was induced by a laser
beam of argon Spectra Physic Laser Stability $2017$ at a wavelength
of $514.5$nm. The beam was focused using a long-frontal x$50$
objective (numerical aperture $0.5$) on an area of about
$3\mu$m$^2$. The laser power on the sample was approximately $20$mW
and the acquisition time was 1 minute. The spectral resolution was
$3$cm$^{-1}$, with a wavenumber precision better than $1$cm$^{-1}$.
The Raman spectra were collected at five points, which were randomly
distributed throughout the mixture. The average of all spectra was
considered as the Raman spectrum of the corresponding mixture for
the considered temperature value and aging time. Raman spectra were
collected $2$ minutes after the beginning of the experiment for
various temperatures ranging between $20$°C and $70$°C in order to
determine the influence of temperature on the polymorph
precipitation. Moreover for each temperature, Raman spectra were
collected at regular time intervals for monitoring phase
transformation. Finally, a total of $N=37$ Raman spectra of $L=477$
wavelengths have been obtained.

\subsection{Data preprocessing}
\begin{figure}[h!]
\centering{
\includegraphics[width=0.5\textwidth]{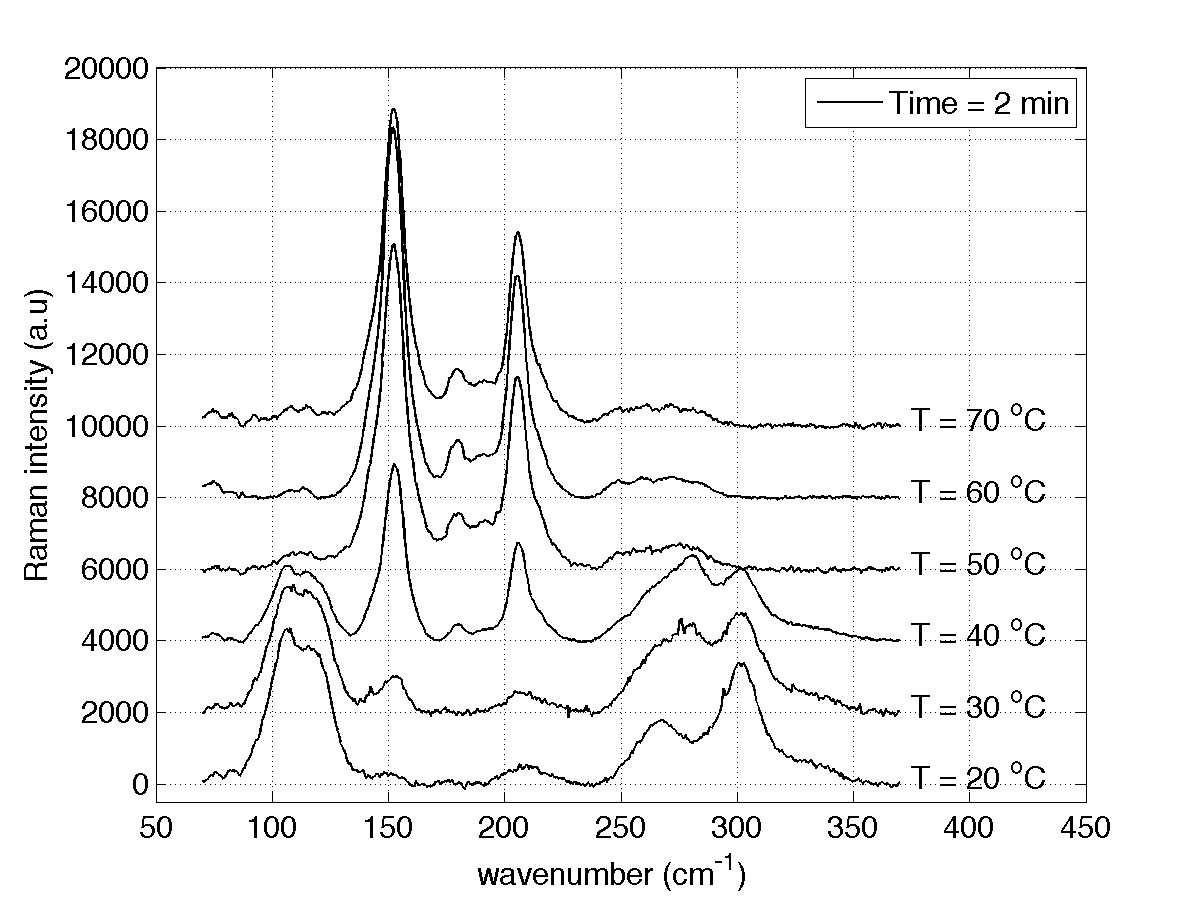}
\caption{Mixture spectra at the beginning of the phase
transformation.} \label{fig:mixturespectra}}
\end{figure}

The Raman spectra of the polymorph mixture are firstly processed
using a background removal approach proposed in
\cite{Mazet2005chem}. In this method, the baseline is represented by
a polynomial whose parameters are estimated by minimizing a
truncated quadratic cost function. This method requires the
specification of the polynomial order and the threshold of the
quadratic cost function truncation. This method was applied for each
spectrum separately with a fifth order polynomial and a threshold
chosen by trial and error. Figure \ref{fig:mixturespectra} shows the
Raman spectra at the beginning of the phase transformation step,
after background removal.

\subsection{Polymorph mixture separation under non-negativity and full additivity contraints}
\begin{figure}[h!]
\centering{
\includegraphics[width=0.5\textwidth]{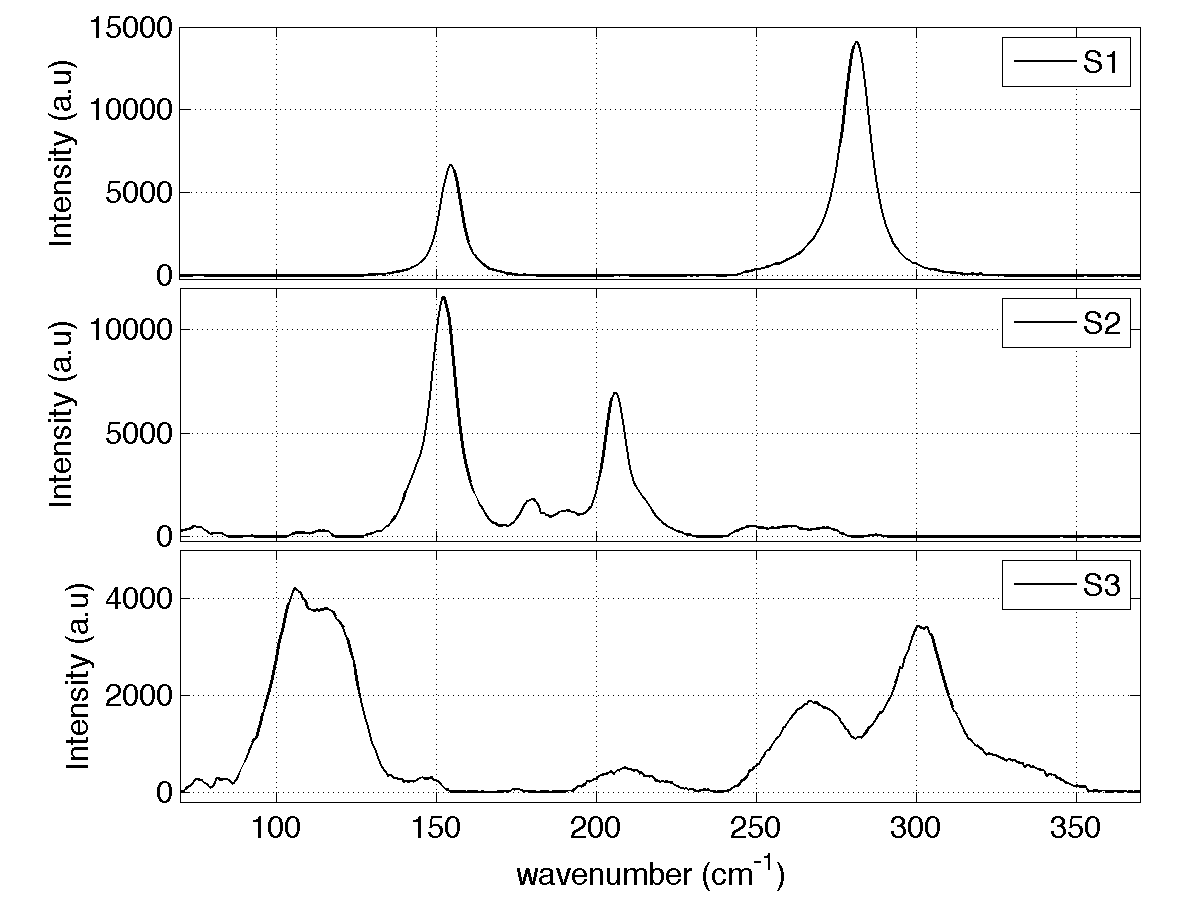}
\caption{Estimated sources.} \label{fig:purespectra}}
\end{figure}

The number of sources to be recovered is fixed to $\nbsou = 3$
according to the prior knowledge on the mixture composition. The
iteration number is fixed to $1000$ iterations where the first $200$
samples are discarded since they correspond to the burn-in period of
the Gibbs sampler. Figure \ref{fig:purespectra} illustrates the
estimated spectra using the proposed approach incorporating the
non-negativity and the full additivitty constraints.

From a spectroscopic point of view and according to the positions of
the vibrational peaks, the identification of the three components is
very easy: the first source corresponds to Calcite, the second
spectrum to Aragonite and the third one to Vaterite. A measure of
the dissimilarity between the estimated spectra and the measured
pure spectra of the three components gives $4.56$\% for Calcite,
$0.65$\% for Aragonite and $4.76$\% for Vaterite. These results show
that the proposed method can be applied successfully without
imposing any prior information on the shape of the pure spectra.

\begin{figure}[h!]
\centering{
\includegraphics[width=0.5\textwidth]{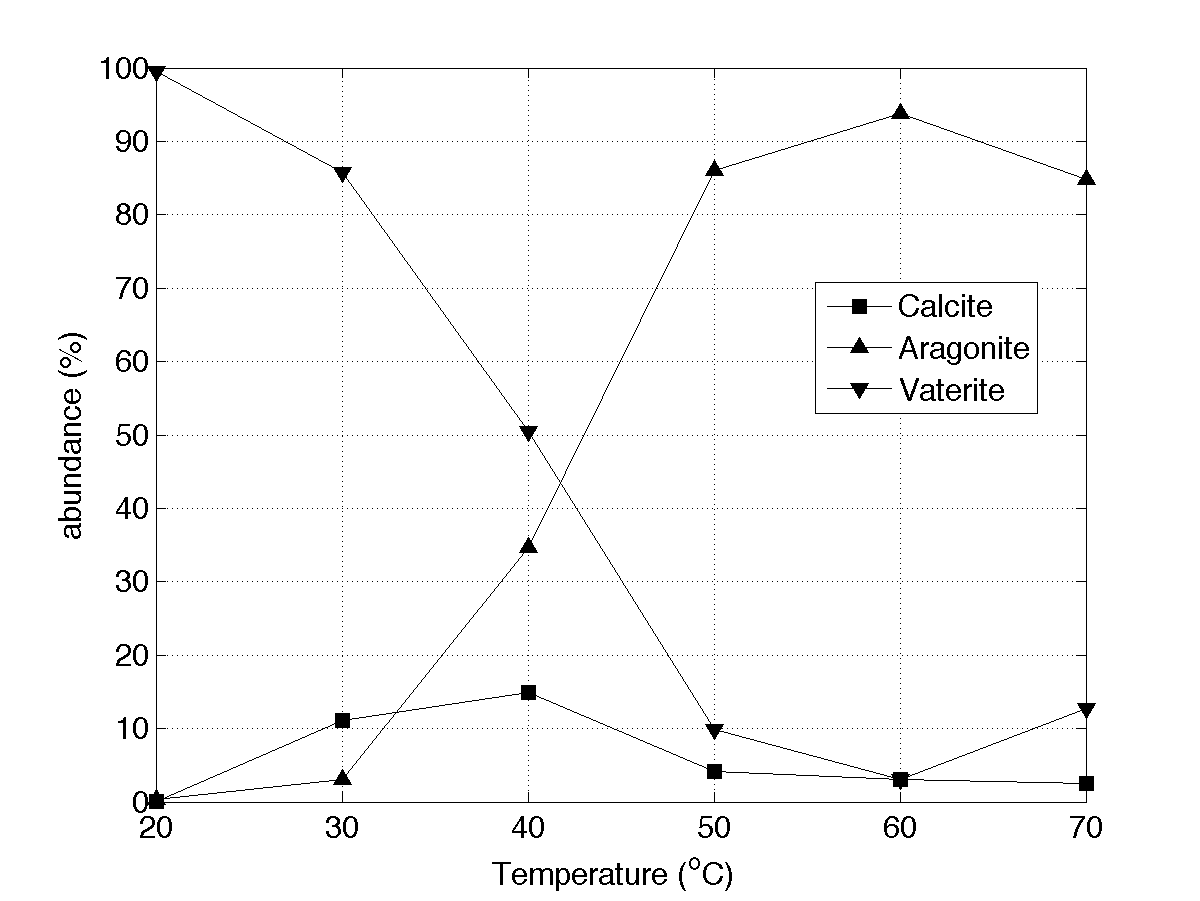}
\caption{Three component abundances at the beginning of the phase
transformation for different temperature values.}
\label{fig:abundances_2min}}
\end{figure}

\begin{figure}[h!]
\centering{
\includegraphics[width=0.5\textwidth]{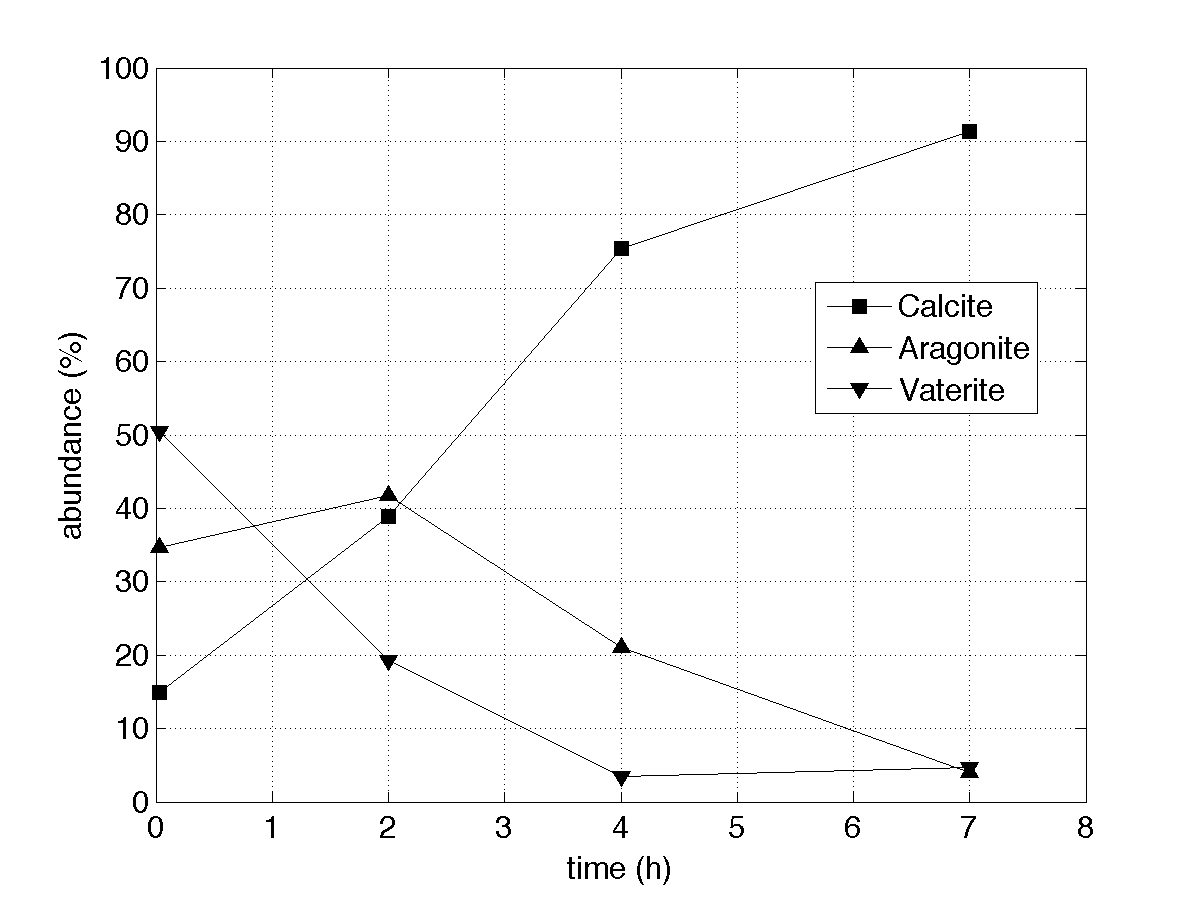}
\caption{Evolution of the three component abundances for $T=40$°C.}
\label{fig:abundancesT40}}
\end{figure}

The evolution of the polymorph proportions versus temperature is
shown in Fig. \ref{fig:abundances_2min}. Pure Vaterite is observed
at $20$°C and a quite pure Aragonite is obtained at $60$°C. However,
between $20$°C and $60$°C ternary mixtures are observed. The content
of Calcite is maximal at $40$°C. Let us now consider the phase
transformation evolution at this temperature value. The
concentration profile versus precipitation time at 40°C is reported
in figure \ref{fig:abundancesT40}. At the beginning of the phase
transformation ($2$ minutes), the ternary mixture is composed of
$50$\% Vaterite, $35$\% Aragonite and $15$\% Calcite. After $2$
hours, the Vaterite is transformed to Aragonite and Calcite. After
$7$ hours, Vaterite and Aragonite are almost totally transformed to
calcite. So, aging time promotes the formation of Calcite which is
in agreement with some results reported in the literature
\cite{Kitamura2002,Dandeu2006}.

\subsection{Polymorph Mixture analysis using other BSS algorithms}
The dataset resulting from this experiment is also used to compare
the performances of standard BSS methods taking into account the
non-negativity constraint and their re-scaled versions ensuring the
full additivity constraint. Table \ref{tab:comparisonexpe}
summarizes the performances of the considered separation algorithms
in terms of normalized mean square errors, dissimilarity measures
and computation times. It can be noticed that the proposed approach
provides source estimates with better accuracy than the other
methods. In addition to the good estimation quality, the second
advantage of the proposed method is its ability to scale the sources
during the estimation algorithm. Thus it does not require any
post-processing of the estimation results. However, as previously
highlighted, the price to pay for having such results is the
computational times required by the proposed MCMC-based estimation
method.

\begin{table}[h!]
\renewcommand{\arraystretch}{1.8}
\centering \caption{Estimation performance for different BSS
algorithms on real spectroscopic data.}
\begin{tabular}{lccc}
\hline
\multicolumn{1}{c}{}   &  $\mathrm{NMSE} (\MATsou)$ & $\mathrm{Diss}(\MATsou)$ & Time (sec)\\
\hline
 Proposed approach     & 0.0072 & 3.34 & 146 \\
 BPSS                  & 0.0118 & 4.87 & 205 \\
 re-scaled-BPSS         & 0.0124 & 4.87 & 205 \\
 NNICA                   & 0.1007 & 11.82 & 29 \\
 re-scaled NNICA          & 0.3996 & 11.82 & 29 \\
 NMF                   & 0.0093 & 4.25 & 26 \\
 re-scaled NMF          & 0.0109 & 4.25 & 26 \\
\hline
\end{tabular}
\label{tab:comparisonexpe}
\end{table}

\section{Conclusion} \label{sec:conclusions}
This paper studied Bayesian algorithms for separating linear
mixtures of spectral sources under non-negativity and full
additivity constraints. These two constraints are required in some
applications such as hyperspectral imaging and spectroscopy to get
meaningful solutions.  A hierarchical Bayesian model was defined
based on priors ensuring the fulfillment of the constraints.
Estimation of the sources as well as the mixing coefficients was
then performed by using samples distributed according to the joint
posterior distribution of the unknown model parameters. A Gibbs
sampler strategy was proposed to generate samples distributed
according to the posterior of interest. The generated samples were
then used to estimate the unknown parameters. The performance of the
algorithm was first illustrated by means of simulations conducted on
synthetic signals. The application to the separation of chemical
mixtures resulting from Raman spectroscopy was finally investigated.
The proposed Bayesian algorithm provided very promising results for
this application. Particularly, when the computational times is not
a study constraint, the proposed method clearly outperforms other
standard NMF techniques, which can give approximative solutions
faster. Perspectives include the development of a similar
methodology for unmixing hyperspectral images. Some results were
already obtained for the unmixing of known sources. However, the
joint estimation of the mixing coefficients (abundances) and the
sources (endmembers) is a still an open and challenging problem.

\section*{Acknowledgments}
\label{sec:Ack} This work was supported by the French CNRS/GDR-ISIS.
The authors would like to thank the anonymous referees and the
associate editor for their constructive comments improving the
quality of the paper. They also would like to thank Prof. Petar M.
Djuri\'c from Stony Brook University for helping them to fix the
English grammar.

\bibliographystyle{IEEEtran}
\bibliography{biblio}
\end{document}